\documentclass[%
 reprint,
 amsmath,amssymb,
 aps,
pre,
floatfix
]{revtex4-2}
\usepackage{amsmath}
\usepackage{graphicx}
\usepackage{dcolumn}
\usepackage{bm}
\usepackage{hyperref}
\hypersetup{
    colorlinks=true,
    linkcolor=blue,
    citecolor=blue,
    filecolor=blue,      
    urlcolor=blue
}
\usepackage[caption=false]{subfig}
\usepackage{textcomp,upgreek}
\renewcommand{\vec}[1]{\bm{#1}}

\newcommand{\ncr}{n}
\newcommand{\La}{_\text{L}} 
\newcommand{\e}{\mathrm{e}} 
\newcommand{\dd}{\mathrm{d}} 

\newcommand\mean[1]{\langle #1 \rangle}

\begin{document}


\title{Cosmic rays and random magnetic traps}

\author{Devika Tharakkal}
 \affiliation{School of Mathematics, Statistics and Physics, Newcastle University, Newcastle upon Tyne, NE1 7RU, UK.}
\author{Andrew P.~Snodin}
 \affiliation{School of Mathematics, Statistics and Physics, Newcastle University, Newcastle upon Tyne, NE1 7RU, UK.}

\author{Graeme R.~Sarson}
 \affiliation{School of Mathematics, Statistics and Physics, Newcastle University, Newcastle upon Tyne, NE1 7RU, UK.}
\author{Anvar Shukurov}
 \affiliation{School of Mathematics, Statistics and Physics, Newcastle University, Newcastle upon Tyne, NE1 7RU, UK.}
\date{\today}

\begin{abstract}
The spatial distribution of cosmic ray (CR) particles in the interstellar medium (ISM) is of major importance in radio astronomy,  where its knowledge is essential for the interpretation of observations, and in theoretical astrophysics, where CR contribute to the structure and dynamics of the ISM. Local inhomogeneities in interstellar magnetic field strength and structure can affect the local diffusivity and ensemble dynamics of the cosmic ray particles. Magnetic traps (regions between magnetic mirrors located on the same magnetic line) can lead to especially strong and persistent features in the CR spatial distribution. Using test particle simulations, we study the spatial distribution of an ensemble of CR particles (both protons and electrons) in  various magnetic field configurations, from an idealized axisymmetric trap to those that emerge in intermittent (dynamo-generated) random magnetic fields. We demonstrate that both the inhomogeneity in the CR sources and the energy losses by the CR particles can lead to persistent local inhomogeneities in the CR distribution and that the protons and electrons have different spatial distributions. Our results can have profound implications for the interpretation of the synchrotron emission from astronomical objects, and in particular its random fluctuations.
\end{abstract}
\maketitle

\section{Introduction}\label{I}

Cosmic rays (CR) are widely recognized to play a vital role in galaxies and serve as an important observational probe, especially in radio astronomy \citep{Longair2011,Jaffe2019}. 
They contribute to the interstellar medium (ISM) pressure equally with the gas, magnetic fields and turbulent flows 
\citep{ShSu21},
making them an important factor affecting both the global distribution of the interstellar gas (such as the disc thickness) and the local structure of the ISM. Furthermore, they can play a crucial role in driving systematic gas outflows (winds and fountains)
from spiral galaxies \citep{I75,BMV91,2008ApJ...674..258E}, and thus contribute to the regulation of star formation and galactic evolution
\citep{MoBoWh10}. The intensity of synchrotron emission depends on both the number density of CR electrons and the magnetic field, so the interpretation of observations of the synchrotron emission requires detailed understanding of their spatial distributions
\citep{G90}.

While the microphysics of particle propagation in the galactic magnetic fields has been studied extensively, a clearer understanding of the ensemble dynamics and local spatial distribution is required to obtain a comprehensive picture of cosmic rays in the ISM. However, the spatial distribution of CR particles at scales comparable to or smaller than the scale of the interstellar turbulence has received rather little attention, as most studies have been focused on CR diffusion and confinement in galaxies. Most of the interpretations of synchrotron radio emission rely on the assumption that the CR and magnetic field energy densities or pressures are equal to each other at any position (the equipartition assumption) \citep[][and references therein]{Seta2019}.

The local magnetic field structure  controls the particle dynamics, and one of the most interesting spatial features affecting cosmic ray propagation is the magnetic trap \citep{Klepach1995,Silsbee2018,Isenberg1979}: a field structure formed between 
{magnetic mirrors (converging magnetic lines) lying on the same magnetic line.} Such magnetic traps can be visualized as a region of a relatively weak field flanked by two regions of a stronger field.
The propagation and distribution of charged particles (either relativistic or non-relativistic) in complex magnetic fields is a central theme in plasma physics and astrophysics, with deep connections to the magnetohydrodynamics of complex fluids. 
Our approach to deriving the particle number density from simulations of individual particle trajectories, and discussion of the associated statistical challenges and biases, may be useful in broader contexts of general physics and biophysics.

A particle of appropriate energy and pitch angle, moving along the field line with its magnetic moment conserved adiabatically, bounces between the mirror points when inside a magnetic trap. 
A detailed description of the principle behind magnetic mirroring is discussed under the trajectories of individual charged particles in basic plasma physics texts \citep{nicholson1983introduction,Kulsrud2005}. The particles can escape from such traps due to field line wandering \citep{1969ApJ...155..777J}, pitch angle scattering from magnetic field variations at scales smaller than the trap (for example, for traps in random fields), and stochastic scattering \citep{Chirikov1960}. The trapping time scales and scattering of a single particle from an ideal magnetic trap  have been studied extensively \citep{Balebanov1967}. 
For an ideal axisymmetric magnetic bottle (i.e., a trap formed between two magnetic mirrors),
the particle is expected to be trapped for long times once the initial pitch angle satisfies the trapping conditions. However, deviation from perfect 
magnetic moment conservation can cause stochastic scattering from the field lines \citep{Dalena2012,Lopez-Barquero2015}, and this can in turn prevent a single particle from being trapped long enough to produce a significant inhomogeneity in the overall particle distribution.
Such local features in inhomogeneous magnetic fields can also affect the local cosmic ray diffusivity, depending on the Larmor radius of the particles and the length scale of the field variations \citep{Snodin2016,Klepach1995,Chandran2000,Chandran2001}.

For an ensemble of particles with a certain (e.g., isotropic) distribution in the angle between their velocities and the local magnetic field direction (the pitch angle), understanding the existence and evolution of a  magnetic trap signature in the spatial distribution of CR in the ISM is a challenging task. 
The confinement of an isotropic plasma in a magnetic trap has been extensively studied in application to various environments from laboratory plasmas to astrophysical systems. However, most such studies focus on the behavior of individual particles rather than their statistical ensemble.  For example, \citet{Chirikov1960} discusses the long-term trapping of a single particle in an idealized trap and the role of stochastic particle scattering. 

Studies on magnetic mirror machines \citep[and references therein]{2016ippc.book.....C} discuss the range in momentum space under which plasma can be trapped in a magnetic trap, and state that isotropic plasma cannot be trapped indefinitely.
Most of these works study the loss parameters of the plasma and the injection conditions required to sustain the plasma trapping for the feasibility of these mirror machines. 
The problem we try to address builds on this and aims to quantify the difference in number densities resulting from magnetic trapping in the context of astrophysical magnetic fields with more realistic injection models.
The problem of containing a statistical ensemble of particles in random magnetic traps has been studied analytically by previous authors. 
The analytical solutions \citep[e.g.][]{1976ApJ...205..900E,1977PhDT........10B,1981ApJ...243.1103S} model the adiabatic focusing of charged particles in an inhomogeneous field by looking at the solutions to Sturm--Liouville operators. 
They study the one-dimensional solutions of the Vlasov equation, where the distribution function depends on the coordinate parallel to the mirror axis, the pitch angle, and time, with the diffusion process modelled as pitch angle scattering. 
They consider the competition between focusing and scattering, and the resulting diffusion function. 
In this study, we take a step back to study the number density variations resulting from mirroring. We do not assume any analytical forms for the focusing, nor do we include pitch angle scattering. Our model is aimed at obtaining the number density distribution arising as a result of the focusing effect of converging field lines.

Among
studies of the propagation of charged particles in weakly inhomogeneous fields we mention \citet{Balebanov1967}, \citet{Ripperda2017} and \citet{Xu2020} as most relevant in the present context. The effect of magnetic structures such as a magnetized molecular cloud, and the corresponding inhomogeneity in the CR particle distribution, is discussed by \citet{Silsbee2018}. Our previous test particle simulations with emphasis on their spatial distribution \citep{Seta2018} addressed the distribution of CR protons in random magnetic fields (both Gaussian and spatially intermittent) and their trapping, and found no correlation between particle distribution and magnetic field strength, thus refuting the equipartition assumption when applied at the scales comparable to or smaller than the correlation scale of the magnetic field.
Moreover, the trapping of the CR particles causes their number density to be larger between the magnetic  mirrors, facilitating an anti-correlation between the CR and magnetic field energy densities.

Apart from trapping between magnetic mirrors, the CR particle distribution can be affected by closed magnetic field lines around elliptic (O-type) magnetic neutral points, which should be abundant in a random magnetic field: such a closed magnetic loop can be either over- or under-populated by the charged particles depending on their sources and pitch-angle scattering (which allows the particles to move across the magnetic field). We do not discuss here the effects of elliptic magnetic null points on the CR distribution, but focus on the trapping of cosmic ray particles between magnetic mirrors.

We explore the spatial particle distribution in a selection of magnetic field configurations (including random ones) and for homogeneous and non-homogeneous CR source distributions, to identify and analyze the effect of magnetic traps on both protons and electrons. The latter lose energy to synchrotron and inverse Compton emissions, which affects their spatial distribution and enhances the trapping \citep{Baker1964a}.

The text is structured as follows. In Section~\ref{sec:magnetic_fields}, we describe the magnetic field configurations used to study the mirroring in static magnetic fields. Section \ref{sec:TPS} introduces the governing equations and physical processes involved in the propagation of both CR protons and electrons. In section \ref{sec:Numericalsetup} we present the numerical setup, discussing particle injection, boundary conditions and physical scales involved in the simulations. The results are presented in section \ref{sec:Results} for both CR protons and electrons.


\section{\label{sec:magnetic_fields}Particle Trapping and Magnetic Field Models}

The evolution of the distribution of CR particles is studied for two spatial configurations of the particle source. In the first case the particles are drawn from a statistically homogeneous random distribution, and in the second case the particle source is a spherical shell surrounding the trapping region, as appropriate for the ISM given that the particles spread from discrete sources such as supernova remnants. The inhomogeneity of the CR sources is especially important in this context since the Liouville theorem precludes the development of any inhomogeneities in a perfectly statistically-homogeneous system, provided the scale of the magnetic field variation is larger than the Larmor radius of the particles and the scattering of the particles (in particular, the pitch-angle scattering) can be neglected. In fact, the Liouville theorem does not preclude the existence of particles that are trapped for an infinitely long time in a static magnetic field, although the set of the initial positions of such particles in the six-dimensional phase-space has measure zero \citep{Baker1964a}. However, trapping for a finite but arbitrary long time is consistent with the Liouville theorem. When particles are injected through a face of a cubic region $V$ with velocity directions confined to a solid angle $\Delta\omega$, their average number density $\bar{n}$ within $V$ has an upper limit \citep{Baker1964a}
$\bar{n}\leq n_0\Delta\Omega/\Delta\omega\,,$
where $n_0$ is the particle number density at the injector and $\Delta\Omega$ ($\leq\Delta\omega$) is the solid angle subtended by the particle velocities within $V$.

Constraints on the distribution of an ensemble of particles in static magnetic fields implied by the Liouville theorem are relaxed in many realistic situations. In particular, the Liouville theorem does not apply when the magnetic field varies on scales smaller than the Larmor radius. The particle deflections by random magnetic fields at sub-Larmor scales lead to a diffusion term in the Fokker--Planck equation obtained by averaging of the kinetic equation for the particle distribution function over a scale comparable to the particle Larmor radius. The phase-space volume occupied by particles in such a diffusive system can decrease with time, leading to an inhomogeneous distribution.
As mentioned above, another factor affecting the consequences of the Liouville theorem is the fact that CR particle injection is non-homogeneous and/or non-isotropic. The CR particles are not `created' uniformly at any position in the ISM but rather spread from discrete sources \citep{EABA21}. The trapping of CR as they penetrate into molecular clouds is discussed by \citet{Silsbee2018}. Moreover, the particle energy losses, especially strong for the CR electrons, lead to further violation of the conditions of the Liouville theorem \citep{Baker1964a}.

We consider two types of magnetic trap. An idealized, axisymmetric trap introduced in Section~\ref{subsec:danmirror} is used to assess the efficiency of the particle trapping and the sensitivity of the particle distribution in space
to the form of their source, particle energy and, in the case of relativistic electrons, energy losses. Our conclusions are further tested by simulations of particle trapping in a magnetic dipole (Appendix~\ref{aADF}). The case of a more realistic trap is discussed in Section~\ref{DGF}, where we use a realization of a random magnetic field generated by the fluctuation dynamo in a random flow and focus on one of the local regions where the number density of test particles has a strong maximum suggesting efficient trapping. Such magnetic traps can be expected to occur in random magnetic fields, and \citet{Seta2018} have shown that they are equally widespread in both Gaussian random magnetic fields and spatially intermittent, strongly non-Gaussian fields produced by the fluctuation dynamo.

Interstellar random magnetic fields vary on a time scale of order $10^6\,\rm yr$ at the integral scale (of order $100\,\rm pc$), much longer than the inverse Larmor frequency of CR particles in a very wide range of energies. As we show below, the spatial distributions of the CR particles settle into stationary states on the relatively long time scale of order $10^4\,\rm yr$ (which is, however, much shorter than the confinement time of the CR particles in galaxies, $\simeq10^7\,\rm yr$).  Therefore, the time variation of the magnetic field can be neglected, and static (time-independent) magnetic configurations can be used for our purposes.

The dimensionless parameter that controls the particle behavior is the ratio of the Larmor radius $r\La$ to the length-scale of $\vec{B}$, so our arguments and results can be re-scaled straightforwardly to other particle energies and magnetic field strengths and scales.

\subsection{Axisymmetric trap}\label{subsec:danmirror}

\begin{figure}
    \centering
    \includegraphics[width=0.4\textwidth]{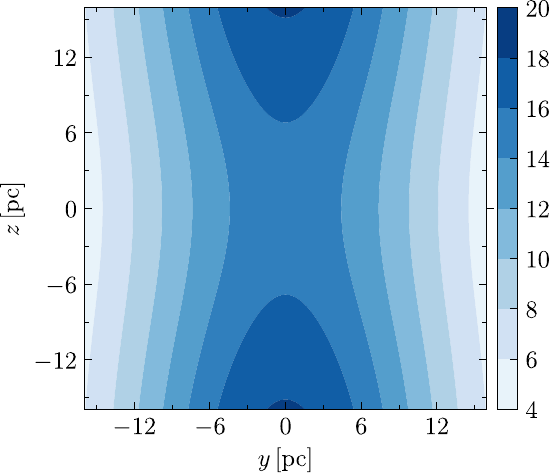}
    \caption{ The cross-section of the axisymmetric magnetic trap of Section~\ref{subsec:danmirror} through the symmetry axis, with the magnetic field strength in $\upmu\rm G$ color-coded.}
    \label{fig:dan2dcross}
\end{figure}

An axially symmetric magnetic trap used to explore general aspects of the CR particle distribution in space has the magnetic field components given in cylindrical coordinates $(r,\phi,z)$ by
\begin{subequations}\label{eq:dan_field}
    \begin{align} 
        B_r &= -{B_0}\, \frac{b }{\ell^2} z f(r)\e^{-z^2/\ell^2}\,, \\
        B_{\phi} &= 0\,, \\
        B_z &= 
        B_0 \left[0.015+ (1 - a r^2)\e^{-r^2/R^2}
            \left(1 - b\e^{-z^2/\ell^2}\right)\right], 
    \end{align}
\end{subequations}
where
\[
f(r) = \frac{R^2}{r} \left[ \e^{-r^2/R^2}(ar^2 + aR^2-1) - aR^2 +1 \right],
\]
and the parameters $R$, $a$, $b$ and $\ell$ control the length scales of the field variation and the positions of the magnetic mirrors. 
(Starting from the chosen analytic form of $B_r$, this field is obtained by requiring solenoidality in axisymmetric cylindrical geometry.)
As illustrated in Fig.~\ref{fig:dan2dcross}, two maxima in the magnetic field strength are located on the $z$-axis, and their separation (i.e., the length of the trap) is controlled by the parameter $\ell$, while the radial scale of $\vec{B}$ depends on $a$ and $R$. Having in mind applications to the ISM of spiral galaxies, we adopt for $\ell$ and $R$ values comparable to the integral scale of the interstellar random magnetic fields,  $\ell=R=16\,\rm pc$
(see Section~\ref{DGF} for the motivation of the scale length choice).

The field structure is further determined by the factors $a = (0.03\, \rm pc)^{-2} $ and $b=0.25$. The positions along the $z$-axis at which particles are reflected depend on the particle energy.
In the simulations presented below, we adopt $B_0=100\,\upmu\rm G$ giving the range $8\leq B\leq 92\,\upmu\rm G$ for the field strength within the computational domain, with the r.m.s.\ field strength of $B_\text{rms}=42\,\upmu\rm G$.

\subsection{Magnetic trap in a random field}\label{DGF}
Magnetic traps occur in virtually any random magnetic field as regions between magnetic mirrors where magnetic lines converge repeatedly towards a magnetic line that serves as the backbone of the trap. \citet{Seta2018} discuss the trapping of charged particles in an intermittent random magnetic field produced by the fluctuation dynamo and a Gaussian random magnetic field with the same power spectrum, and show that both types of random magnetic field produce numerous traps.

\begin{figure}
    \centering
    \includegraphics[width=0.46 \textwidth]{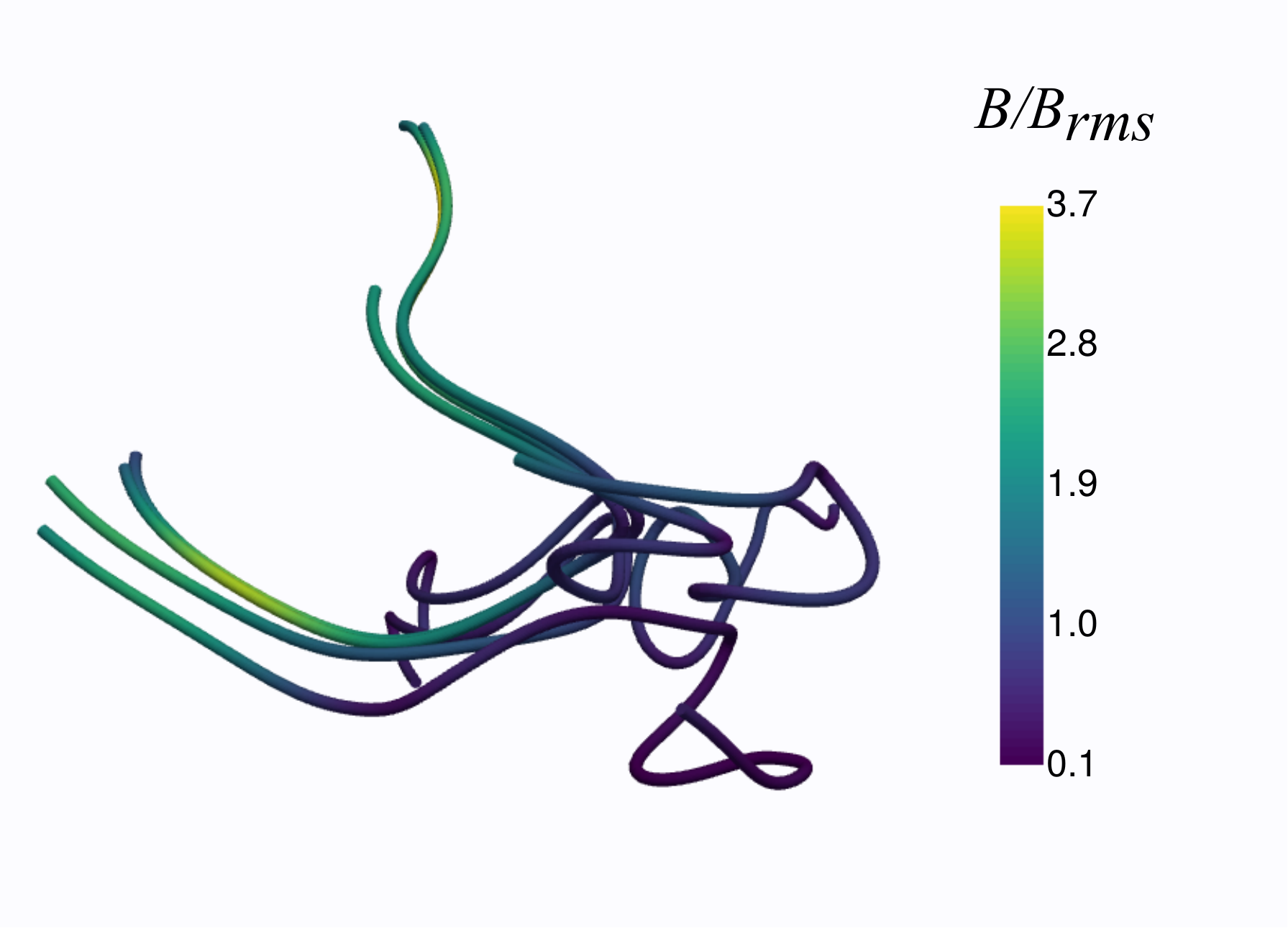}
    \caption{The three-dimensional structure of the magnetic lines of the random magnetic trap of Section~\ref{DGF},
    with the field strength relative to its root-mean-square value color coded.
    }
    \label{fig:bint_iso}
\end{figure}

An example of such a trap in an intermittent magnetic field is shown in Fig.~\ref{fig:bint_iso}. This magnetic field is obtained as a solution of the induction equation with a time-dependent, multi-scale velocity field with chaotic trajectories \citep{WBS07,SSSBW17}.

The dynamo action produces a non-Gaussian random magnetic field represented by magnetic filaments and ribbons even when the velocity field realizations have Gaussian statistics \citep{SS21}. The magnetic structure shown in Fig.~\ref{fig:bint_iso} is from a region where the CR proton distribution has a strong local maximum in the simulations of \citet{Seta2018}, and we discuss here in detail the behaviour of the statistical ensembles of CR protons and electrons in this particular trap, including the long-term evolution of the spatial particle distributions. We note once more that the occurrence of magnetic traps does not rely on the magnetic intermittency: this is a generic feature of random magnetic fields.

The dynamo simulations \citep{SSSBW17} which produced the magnetic structure of Fig.~\ref{fig:bint_iso} were performed on a $512^3$ grid in a periodic box of the dimensionless size (edge length) $2\pi$ corresponding to the physical size comparable to the integral scale of the interstellar turbulence, 100\,pc {\footnote{The magnetic field data are available at \url{https://doi.org/10.5281/zenodo.4382442} as file \texttt{B\_Rm3182.nc}.}}. In our simulations, we use the same unit length, $L = 100 \, {\rm pc}/(2\pi) = 16\,\rm{pc}$. The field strength within the computational domain ranges from  $0.025\,\upmu\rm G$ to $89 \,\upmu\rm G$ while the r.m.s.\ field strength is $B_\text{rms}=5\,\upmu\rm G$.
(Here we might comment on this variability of magnetic field strength in our simulations. The turbulent motions driven by supernovae in the ISM are highly compressible in nature, and the resulting fluctuation dynamos can generate magnetic fields with such a wide range of field strengths \citep{Gent2021}.)

\section{Basic equations and test particle simulations}\label{sec:TPS}
When the particle energy losses and gains can be neglected, the dimensionless equations of motion of a particle with electric charge $q$  ($q=-e$ for an electron)
and rest mass $m$ moving at velocity $\vec{v}$ in a magnetic field $\vec{B}$ have the form

\begin{align}\label{eq:lorentzforce}
\frac{\dd\vec{p}'}{\dd t'} &= \frac{\alpha}{\gamma} \vec{p}' \times \vec{B}'\,,\\
\label{lorentz2}
    \frac{\dd\vec{x}'}{\dd t'} &= \frac{\beta}{\gamma} \vec{p}'\,,
\end{align}
where $\vec{x'} = \vec{x}/L$, $\vec {p'} = \vec {p}/mc$, $t' = t/t_0$ and $\vec{B}'=\vec{B}/B_0$ are the dimensionless coordinates, momentum, time and magnetic field, respectively, and
\begin{equation*}
    \alpha = \frac{qB_0t_0}{mc}\,,
    \qquad
    \beta = \frac{ct_0}{L}\,.
\end{equation*}
Here $\vec p$ is the relativistic momentum and the associated Lorentz factor is $\gamma =E/(mc^2)= \sqrt{1+(p/mc)^2}$ with $E$ the particle energy, and $p=|\vec{p}|$. 

Charged particles gyrate around a magnetic line under the influence of the Lorentz force, and its motion is characterized by the Larmor radius, frequency and the angle $\theta$ between the magnetic field and velocity vectors (the pitch angle) \citep{Kulsrud2005},
\begin{equation}\label{eq:Larmor_radius}  
    r_\text{L} = \frac{\gamma m c v}{ |q| B} =\frac{E}{|q|B}\,,
\quad    \omega_\text{L} = \frac{c}{r_\text{L}},
\quad    \cos \theta =\frac{ \vec{v} \cdot \vec{B}}{vB}\,,
\end{equation}
where $v=|\vec{v}|$ and $B=|\vec{B}|$.
The time unit corresponding to the unit length $L=16\,\rm pc$ is $t_0 = 2 \pi L/c\approx 300\,\rm yr$. A suitable unit for the interstellar magnetic field strength is $B_0=5\,\upmu\rm G$, and the particle energy with the Larmor radius $L$ in magnetic field $B_0$ is about   $E_0=10^6\,{\rm GeV}(r_\text{L}/1\,\rm pc)(B_0/1\,\upmu\rm G)\simeq10^8\,\rm GeV$.

Equation~~\eqref{eq:lorentzforce}  is applicable to CR protons since their energy losses can be neglected. The energy losses of CR electrons to synchrotron emission and inverse Compton scattering are not negligible and the electron energy evolves as \citep{Longair2011}
\begin{equation}\label{eq:synch_loss}
\frac{\dd E}{\dd t} = -\kappa  E^2\,,
\end{equation}
where
\[
\kappa = \frac{1}{1.2\times10^{10}\,\rm yr\,GeV}
\left[\left(\frac{B}{1\,\upmu\rm G}\right)^2+\frac{w}{0.25\,\rm eV\,cm^{-3}}\right]\,,
\]
with $w$ the energy density of the ambient radiation.
(We only include losses to the synchrotron emission in our simulations.) As a result, the electron Lorentz factor in Eqs~\eqref{eq:lorentzforce} and \eqref{lorentz2} decreases with time,
and the rate of the decrease varies with the magnetic field strength along the particle trajectory. We note that the electron energy losses due to the inverse Compton scattering from the present-day cosmic microwave background are equivalent to those in a magnetic field of $3.2\,\upmu\rm G$ in strength, weaker than the interstellar magnetic fields in nearby galaxies. 
Exact theoretical formulation requires the synchrotron loss coefficient to depend on the pitch angle.
Here we make use of the isotropy of the injected particles, 
to effectively average over the angular dependencies. 
The inclusion of pitch angle would enhance the energy losses as particles near the mirroring regions. 
Hence our model will give a conservative lower limit to the effect energy losses have on mirroring, and to the resultant inhomogeneity in the particle distribution.

Magnetic mirroring is associated with the conservation of the particle magnetic moment,
\begin{equation}\label{eq:magmoment}
    \mu = \frac{ m v_{\perp}^2}{2B}\,,
\end{equation}
an adiabatic invariant (here $v_\perp=v\sin\theta$ is the particle speed perpendicular to $\vec{B}$), which changes only slightly if the particle propagates through a weakly inhomogeneous field,
i.e., if the magnetic field only varies at scales much larger than the particle's Larmor radius.
To be reflected at a magnetic mirror, the particle has to have a sufficiently large pitch angle: particles that travel from a region with magnetic field strength $B$ towards a mirror with magnetic field strength $B_\text{m}$ avoid the reflection if they are within the loss cone $\theta<\theta_\text{m}$, where $\sin^2\theta_\text{m}=B/B_\text{m}$.

Magnetic field variations at scales smaller than $r_\text{L}$ cause the particles to scatter off the guiding field line \citep{Dalena2012}. Interstellar magnetic fields have a wide range of scales extending down to $10^8\,\rm cm$ \citep{ARS95,CL10}, 
and variations of the magnetic moment play a role in the CR propagation \citep{Lopez-Barquero2015}.

\subsection{\label{sec:Numericalsetup}Numerical implementation}
The particle trajectories are integrated numerically using the eighth-order Runge-Kutta method (DOP853, using  the Dormand and Prince coefficients) with an adaptive time step \cite{HairerEA93}. We select error parameters for the method so that the particle energy in our simulations is conserved to the eighth significant digit when energy losses are neglected.
In the discussion of isolated traps, either axisymmetric or from the random field, the size of the computational domain is $2L=32\,\rm pc$, 
while the sizes of the magnetic traps that we consider is about $L$; the traps are placed at the center of the domain. 
We also conduct simulations for a full realization of a multi-scale
random field where the computational domain is 100\,pc in size and contains many magnetic traps.

The axisymmetric magnetic field of Section~\ref{subsec:danmirror} can be evaluated at any position of a moving particle. The random magnetic field of Section~\ref{DGF} is specified on a three-dimensional grid with $0.16 \,\rm pc$ spacing, and is interpolated locally to the current particle position using trilinear interpolation. 


\subsection{Particle injection}\label{PI}

As argued in Section~\ref{sec:magnetic_fields}, a realistic modeling of the spatial CR distribution in a magnetic trap requires special attention to the spatial form of their injection region. When particles are injected uniformly, isotropically and continuously, the particle distribution in a magnetic field that only varies at scales exceeding the Larmor radius must remain homogeneous as $t\to\infty$ according to the Liouville theorem. However, CR particles are not injected uniformly but rather have discrete sources (mainly, supernova remnants). Therefore, apart from simulations with a uniform injection of particles, designed to confirm that our results are consistent with the Liouville theorem, we consider the physically more relevant results obtained when the particles are injected in a spherical shell around a magnetic trap with the inner and outer radii of $2L/3$ and $L$, respectively.

In both cases, the particles are introduced with an isotropic distribution of their pitch angles and at random positions within the injection region.

To explore steady-state particle distributions, particles that leave the computational domain via its boundaries, or (in the case of electrons) lose their energy to insignificant values, have to be reintroduced into the system to keep the total number of particles approximately constant as the system evolves. 
We use two alternative approaches for this particle reintroduction. 
In one approach we apply a reflection condition at the boundaries, whereby
the particle velocity is reversed as it crosses the boundary (see Appendix~\ref{aADF} for details). In an alternative approach, we reintroduce particles at a random position within the injection region. In order to assess the effect of the re-injection method on the results, 
in some simulations we do not reintroduce the lost particles, so that the total number of particles decreases with time in those simulations. 

For CR electrons with their energy losses, we also consider how their energy spectrum evolves as they propagate, and we consider two types of injection energy spectrum: one where the particles are all injected at the same energy, and one with a power-law injection energy spectrum.

In the former case, all the electrons are injected with same energy $E=E_{\rm max}$ chosen such that their Larmor radius, corresponding to the r.m.s.\ magnetic field in the trap, is comparable to the trap size. 
We also specify a minimum energy $E_{\rm min}$: when a particle energy decreases below $E_{\rm min}$, the particle is removed and re-injected with the energy  $E=E_{\rm max}$. The magnitude of $E_{\rm min}$ is selected to avoid the particle Larmor radius based on the r.m.s.\ field strength decreasing below $0.1$ of that corresponding to $E_{\rm max}$, i.e.,\ $E_\text{min}=E_\text{max}/10$.

For analysis, we bin the particles into $M$ energy intervals of equal width, 
$\Delta E =(E_{\rm max}- E_{\rm min})/M$,and the results are presented with the particle energy $E$ referring to the bin center.

In the case of the spectral energy injection, the energy of an injected particle is drawn at random from the probability distribution proportional to $E^{-s}$ in $E_{\rm min}<E<E_{\rm max}$ to obtain the injection spectrum with the spectral index $s$; we use $s=-3/2$. The values of $E_{\rm max}$ and $E_{\rm min}$ are determined as above and the particle position is evolved until its energy reduces to $E_{\rm min}$.
For the analysis of the spatial particle distribution, they are binned into $M$ unequal energy intervals, with the $i$-th interval ($1\leq i\leq M$) of width $\Delta E_i = N E_i^s/M$, to obtain similar numbers of particles in each energy bin (here $N$ is the total number of particles in the simulation). The energy $[E_i(E_i+\Delta E_i)]^{1/2}$ is used to represent the particles in the energy range $E_i<E<E_i+\Delta E_i$.
Unless stated otherwise, these parameters were chosen as
$(N, E_\text{max}, E_\text{min}, M)=(421875, 10^6\,{\rm GeV}, 10^5\,{\rm GeV},5)$ with $N$ kept a constant in each simulation,
but some results are obtained with three and five times larger values of $N$ for both protons and electrons, and in some cases we do not re-inject particles: then $N$ decreases with time.

\subsection{The number density of particles}\label{NDoaEoP}

For each particle in an ensemble, we computed its trajectory and sampled its position at times separated by an interval $T$ specified below. In the case of electrons, the energy bin to which the particle belongs was also recorded. The magnitude of $T$ was chosen to ensure that the magnetic fields at the particle positions at times $t$ and $t+T$ are sufficiently different; in practice, the two positions are typically more than two Larmor radii apart (in the case of a random magnetic field, the separation of those positions  could be chosen to exceed the integral scale of the field). To ensure that the positions obtained for different particles are compatible, the sampling time interval $T$ in physical units was chosen
to be equal for all particles, $T=516\,\rm yr$, with or without energy losses. Thus, the set of particle positions at any given discrete time $jT$, with integer $j$, can be considered as a snapshot of the spatial distribution of a large number of particles launched simultaneously.
The particle positions were mapped into a cubic grid with $75^3$ volume elements within the computational domain (so the size of a volume element is about $(0.4\,{\rm pc})^3$ ).
The particle number density $n$ at a position $\vec{x}$ was obtained by dividing the total number of particles in the grid element containing $\vec{x}$ by its volume, and then smoothing with a Gaussian kernel (half-width of $0.85 \,\rm pc$). 
The number of particles involved in the simulation was $75^3$, with 100 position values recorded for each of them. The effective total number of particles involved in most of the calculations of the particle number density is $100\times75^3\approx4\times10^7$, and the mean number of particles per the grid volume element is 100. As discussed below, the total number of particles involved is sufficiently large to justify our conclusions.

The number density of particles obtained through the sampling of their trajectories in a finite domain is subject to a bias discussed in Appendix~\ref{aADF} since longer trajectories contribute more strongly to the resulting value of $\ncr$ than shorter trajectories. This effect is pronounced for regular magnetic field configurations where the length of magnetic lines that fit into a finite simulation domain can vary significantly from one line to another. 
As explained there, we compensate this bias by using the reflecting boundary conditions where the particle velocity is reversed when it reaches
the boundary. In the case of a random magnetic field, most magnetic lines have similar length when the computational domain is big enough.
Therefore, useful results can also be obtained using the alternative boundary conditions, whereby particles that leave the domain are re-injected at random interior positions within the injection region.

Populating the magnetic field lines with a sufficient number of particles to obtain statistically meaningful results for the particle number density is challenging, especially in a random magnetic field and when the injection is inhomogeneous. A delicate task here is to obtain enough magnetic field lines connected to the inhomogeneous source so that particles can propagate to the magnetic trap in the interior.
In the case of a simple field structure, its larger scale of variation renders this task easier, compared to the case of the random field which lacks a large-scale mean field. In order to study the effect of inhomogeneous injection on mirroring structures in random fields, we define a spherical shell around the trap in such a way that there are field lines connecting both these regions. 
(The spherical shell region does not represent any specific source or sources, but rather represents a `bath' of sources, external to the internal region being studied.)
Appropriate radii of the spherical-shell injection region were adopted after a few trials. 
The spherical geometry of the source allows for the maximum number of magnetic field lines connecting the source and the trap.
We expect the results to be similar in the case of other inhomogeneous source conditions, if enough particles are populated and propagated along the field lines connecting the trap and the source.
To demonstrate the consistency of the results obtained, 
we also consider a test case for the isolated random trap with a point source in the middle of the box,$(x,y,z)=(0,0,0) \,\rm pc$.

\section{\label{sec:Results}Results}

\begin{figure*}
    \centering
\includegraphics[width=0.9\textwidth]{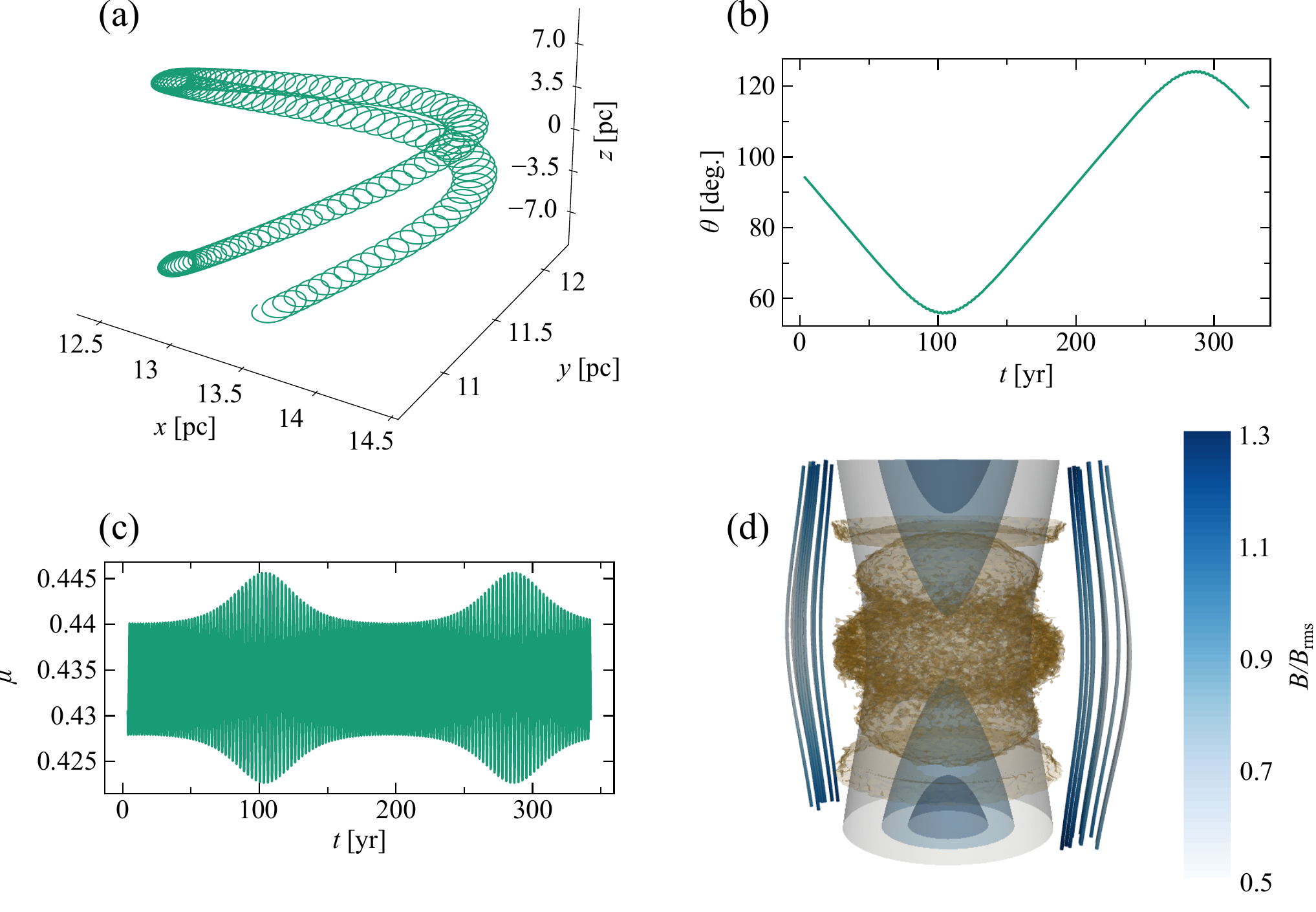}
    \caption{A proton with $r\La= 0.24\, \rm pc$ trapped in the axisymmetric magnetic bottle of Section~\ref{subsec:danmirror}:  \textbf{(a)}~the trajectory; the evolution of \textbf{(b)}~the pitch angle, and \textbf{(c)}~the magnetic moment; \textbf{(d)}~the isosurface of the particle number density $\ncr/\langle\ncr\rangle=0.5$ (brown) at $t=1.4\times10^5$ for a spherical shell injection with the magnetic field lines and isosurfaces $|\vec{B}|/B_\text{rms}= 1.4,1.8$ and 2.2 (blue, with $B_\text{rms}$ the root-mean-square field strength).
}
    \label{fig:dan_mirror_traj}
\end{figure*}

The degree of inhomogeneity in the particle distribution is characterized by
\begin{equation} \label{eq:homogeneity}
\Phi = \frac{  \langle \ncr \rangle^2}{ \langle \ncr^2 \rangle},
\end{equation}
where the angular brackets denote the volume average.
For a perfectly homogeneous system, $\Phi=1$. The smaller is $\Phi$, the stronger is the inhomogeneity; in the extreme case of isolated, uniform clouds with sharp boundaries, $\Phi$ represents their fractional volume.
The magnitude of $\ncr$ depends on the number of particles involved in the simulation and the sampling rate of their trajectories, and can be scaled to any desired value (e.g., $10^{-9}\,{\rm particles}\,{\rm cm}^{-3}$ for the average number density of the Galactic CR, versus the mean number of 100 particles per $(0.4\,\rm pc)^3$ in our simulations). We present our results in terms of the relative number density $\ncr/\langle\ncr\rangle$; together with $\Phi$, this quantity is independent of the normalization and characterizes the trap.

\subsection{Proton distribution in the axisymmetric trap}\label{PDiAT}

\begin{figure*}
    \includegraphics[width=0.8\textwidth]{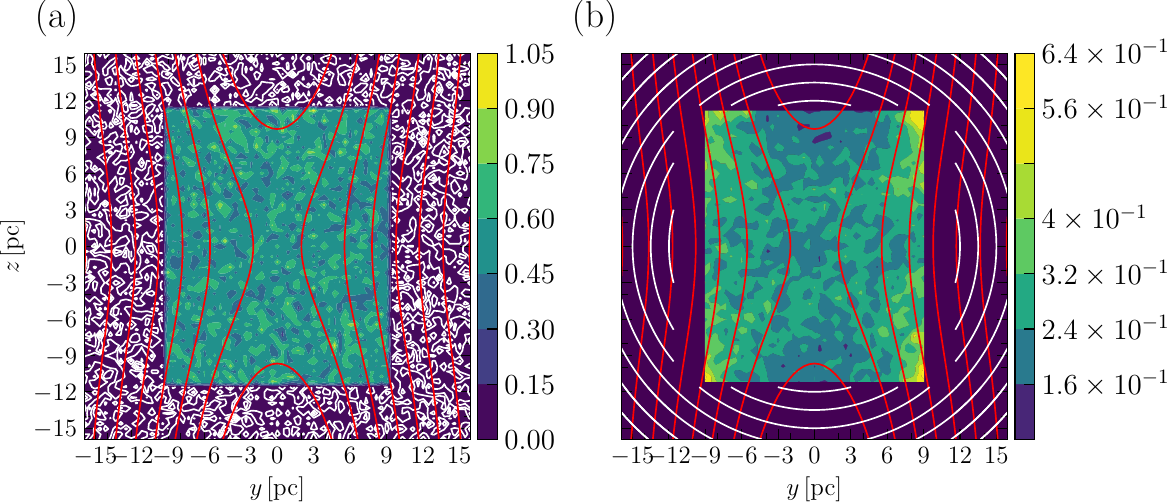}
    \caption{The relative number density  $\ncr/\mean{\ncr}$ of protons
    with $r\La=2.4\,\rm pc$  (color-coded)     in the cross-section of the axisymmetric trap at $x=0$ (integrated over $|x|\leq 6.4\, \rm pc$) at $t=6944\,\rm yr$:
\textbf{(a)}~homogeneous injection with reflecting boundaries and
\textbf{(b)}~injection in a spherical shell with open boundaries and re-injection. The isocontours of $|\vec{B}|$ are shown with red lines and the regions used to compute $\Phi$ (shown in Fig.~\ref{fig:dan_stats} and discussed in the text) are within the inner frames. The white contours outside the inner frames represent the isosurfaces of the relative number density in the injection region.}
    \label{fig:ncr_dan_analytical015}
\end{figure*}

We use the axisymmetric trap of Section~\ref{subsec:danmirror} to clarify and quantify the sensitivity of the results to the simulation parameters (such as the total number of particles, the sampling of their trajectories and the duration of the simulation) as well as to verify the effects of the shape of the injection region on the particle distribution. Since these aspects of the particle behavior are largely independent of the energy losses, we only consider protons in this case. The trajectory of a single proton trapped in the axisymmetric magnetic trap is shown in  Fig.~\ref{fig:dan_mirror_traj}a: at this particle energy, it is reflected at $z=\pm 7.5 \,\rm pc$ and drifts along the azimuth because of the radial variation of the magnetic field strength. Correspondingly, the pitch angle varies periodically and nearly linearly between its extrema with the reflections that occur when $\theta=90^\circ$  (Fig.~\ref{fig:dan_mirror_traj}b), while the magnetic moment only varies by 5\% without any signs of a systematic trend  (Fig.~\ref{fig:dan_mirror_traj}c).
Fig.~\ref{fig:dan_mirror_traj}d shows a three-dimensional perspective view of the axisymmetric trap, with $z$ aligned vertically.  
The different shades of translucent grey to blue show isosurfaces of increasing magnetic field strength.
The gold isosurface shows one value of the resulting number density of cosmic rays. (See also Fig.~\ref{fig:ncr_dan_analytical015}b.)

\begin{figure*}
    \centering
    \subfloat{
    \includegraphics[width=0.487\textwidth]{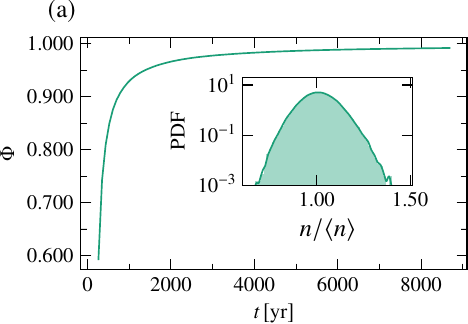}}
    \subfloat{
    \includegraphics[width=0.485\textwidth]{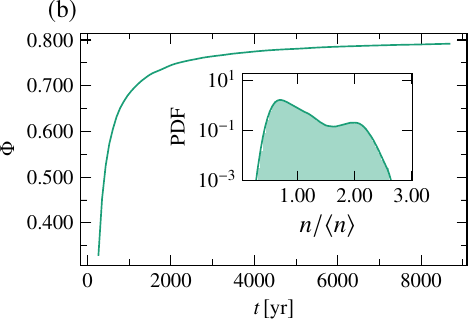}}
    \caption{The evolution of the inhomogeneity measure $\Phi$ for protons
    ($r\La=0.786\,\rm pc$ based on the r.m.s.\ field strength) injected into the axisymmetric trap \textbf{(a)}~homogeneously (with no subsequent re-injection) and \textbf{(b)}~in the spherical shell (with re-injection). The values of $\Phi$ were computed for a cubic region outlined in Fig.~\ref{fig:ncr_dan_analytical015}. The insets show the probability density of $\ncr/\mean{\ncr}$ at 
   $t= 8680\, \rm yr$, obtained as a histogram and using the Gaussian kernel density estimate (solid line). }
    \label{fig:dan_stats}
\end{figure*}

\begin{figure*}
\centering
    \includegraphics[width =0.48\textwidth]{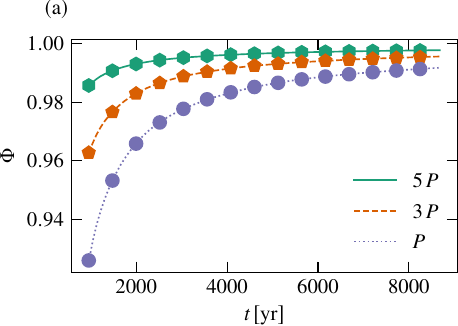}
\hfill
\includegraphics[width=0.48\textwidth]{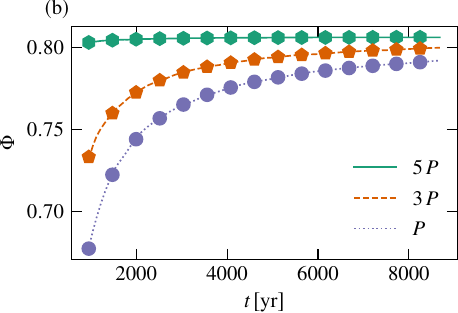}
    \caption{The dependence of inhomogeneity measure $\Phi$ on time for the proton distribution in an axisymmetric trap for various values of the total particle number, $N=P$ (dotted, the same as shown in Fig.~\ref{fig:dan_stats}), $N=3P$ (dashed) and $N=5P$ (solid), where $P=75^3=421875$, for: \textbf{(a)}~homogeneous injection and \textbf{(b)}~injection in a spherical shell.  Note the difference in the values of $\Phi$ at large $t$ between (a) and (b). The computed values of $\Phi$ are shown with continuous curves while the corresponding fits $\Phi=c-dt^{-\gamma}$ described in the text and Table~\ref{tab:conv_case1} are shown with markers.}
    \label{fig:danerrstatconvg}
\end{figure*}

Fig.~\ref{fig:ncr_dan_analytical015} illustrates the effect of the form of the injection region on the relative number density distribution. For a homogeneous and isotropic particle injection (Fig.~\ref{fig:ncr_dan_analytical015}a), the relative density variations of order 20\% are consistent with the relative statistical noise of $(\widetilde{N})^{-1/2}$ expected for the binned data with $\widetilde{N}<100$ particles per spatial bin. 
A slight increase in the relative number density at the ends of the trap
($z=\pm7.5\,\rm pc$) 
is an artifact of the reflecting boundary conditions given that more particles reach the boundary along the magnetic field than across it. The negligible inhomogeneity in the particle distribution in this case is consistent with the implications of the Liouville theorem within statistical errors
(confirmed using numerical experiments with randomly placed particles). We have verified that simulations over longer times and involving a larger number of particles result in weaker density variations in the case of the homogeneous, isotropic particle injection.

In the case of the inhomogeneous injection in a spherical shell (Fig.~\ref{fig:ncr_dan_analytical015}b), the variation in $\ncr/\mean{\ncr}$ is significantly stronger (corresponding to $\Phi\approx0.8$; see below). 
The injection region is visible as the annulus with a higher $\ncr$. 
Our preliminary results indicate that the asymptotic value of $\Phi$ is sensitive to the shape of the injection region, and injection through one face of the cubic domain or at a single point is likely to lead to stronger inhomogeneities.

The variation in the magnetic field strength along magnetic lines is weaker for the lines that pass near to the trap axis (compared to those further off-axis)
since the loss cone is wider for particles that move along a stronger magnetic field.
This explains the reduction of the particle number density near the trap axis visible in Fig.~\ref{fig:ncr_dan_analytical015}b.
(This effect is not evident in Fig.~\ref{fig:ncr_dan_analytical015}a,
as in that case particles are re-injected in this region.)

\begin{table}
\caption{The variation of the inhomogeneity parameter $\Phi$ in the axisymmetric trap with time $t$ and the total number of particles $N$ involved in the simulations, specified as a multiple of $P=75^3=421875$, for the homogeneous and spherical-shell injection regions. Presented are the fitted parameters $c$, $d$ and $\gamma$ of the approximation $\Phi=c-dt^{-\gamma}$,
with $t$ in years. The particle Larmor radius is $r\La=0.786\,\rm pc$, energy losses are neglected.}
    \centering

\begin{ruledtabular}
\begin{tabular}{lccccccc}
    &\multicolumn{3}{c}{Homogeneous}
       & &\multicolumn{3}{c}{Spherical shell}\\ 
        \cline{2-4} \cline{6-8}\\
$N$     &$P$    &$3P$   &$5P$ &  &$P$    &$3P$   &$5P$\\
$c$     &0.998     &0.998     &0.998    &  &0.84   &0.82   &0.81\\
$d$     &1.0    &1.2    &1.3  &  &0.7    &0.9    &1.3\\
$\gamma$&2.0    &2.1    &2.3  &  &1.4    &1.5    &1.9\\
\end{tabular}
\end{ruledtabular}
    \label{tab:conv_case1}
\end{table}
\begin{figure}
    \subfloat{
    \includegraphics[width=0.4\textwidth]{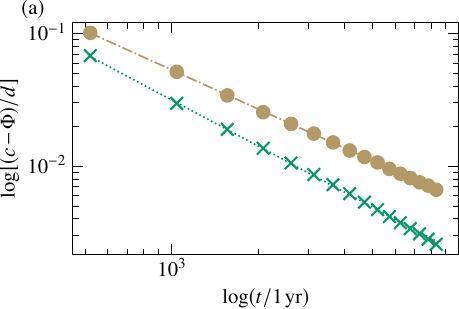}}
    \quad
        \subfloat{\includegraphics[width=0.4\textwidth]{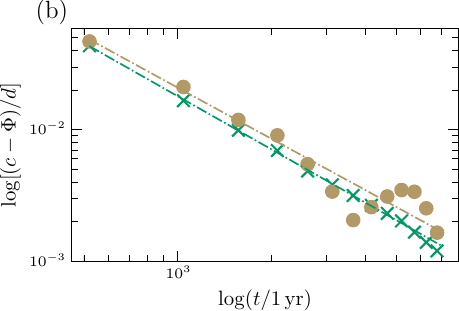}}
    \caption{The quality of the fit $\Phi=c-dt^{-\gamma}$ with parameters from Table \ref{tab:conv_case1} for \textbf{(a)}~the axisymmetric trap and \textbf{(b)}~trap in the random magnetic field (Section~\ref{TiaRF}). The computed values of $\Phi$ are shown with crosses for homogeneous injection and circles for injection in a spherical shell; the fits 
for $t\geq 520 \, \rm yr$ are represented with straight dashed lines.    }
    \label{fig:phi_fit}
\end{figure}
To demonstrate that the inhomogeneity in the particle distribution
is not a short-term transient or an artifact of the limited number of particles $N$ in the simulation, we consider the evolution of the inhomogeneity measure $\Phi$ with time for various values of $N$.
For the time dependence, we calculate $\ncr$ and $\Phi$ as described at the end of Section~\ref{sec:TPS} until a current time $t$ ($0<t\leq10^2 T$), with the particle positions at all earlier times included in the calculation. The smaller is $t$, the smaller is the effective number of particles used to calculate $\ncr$. 

As shown in Figs~\ref{fig:dan_stats} and \ref{fig:danerrstatconvg}, $\Phi$ increases with $t$ (and thus the degree of inhomogeneity decreases) as the number of particles involved in the calculations increases (so that the statistical noise becomes weaker). It is notable, however, that for the homogeneous injection $\Phi$ evidently tends to unity as $t$ increases, whereas $\Phi$ tends to a smaller value in the case of the spherical-shell injection.
The insets in Figs~\ref{fig:dan_stats}a and \ref{fig:dan_stats}b show the probability density 
distribution of the particle number density. 

With the log-linear axes, the log-normal and exponential distributions in Fig.~\ref{fig:dan_stats} are represented by a parabola and straight line, respectively. The insets show that the probability density of $\ncr$ is sensitive to the form of the injection, with an approximately log-normal distribution in the case of the homogeneous injection and a more complicated one, with pronounced high-density features around $\ncr/\mean{\ncr}=2$, in the case of the spherical-shell injection. Moreover, the probability density extends to significantly larger values of $\ncr/\mean{\ncr}$ when the injection is not homogeneous.

\begin{figure*}
\centering
\includegraphics[width=0.84\textwidth]{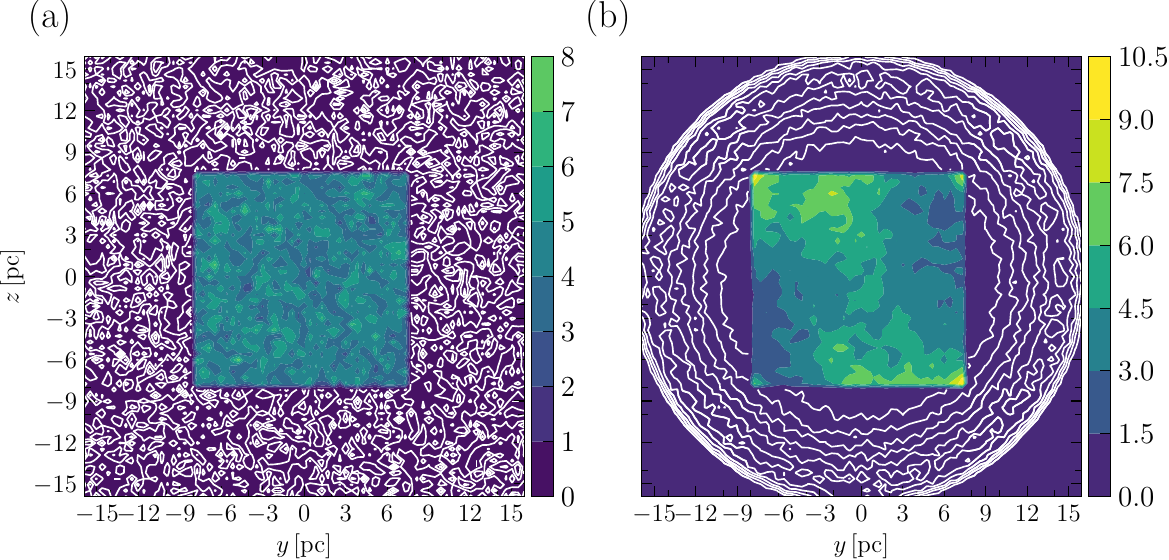}
\caption{As in Fig.~\ref{fig:ncr_dan_analytical015} but for protons with $r\La=1.2 \, \rm{pc}$ in the cross-section of the random trap at $x=0$ (integrated over $|x|\leq 6.4\,\rm pc$) at $t=6944 \,\rm yr$: \textbf{(a)} homogeneous injection with re-injection of the particles lost through the boundaries and \textbf{(b)} injection in a spherical shell with re-injection.
}
\label{fig:ncr_dynamo}
\end{figure*}

\begin{figure}
\subfloat{
\includegraphics[width=0.4\textwidth]{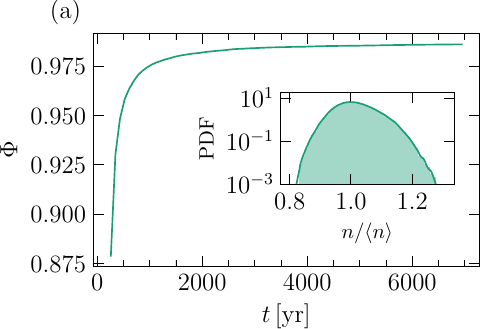}}\\
\subfloat{\includegraphics[width=0.4\textwidth]{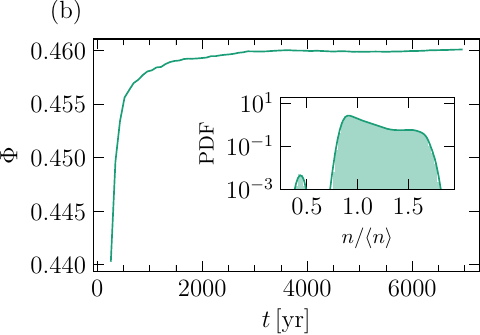}}
\caption{The evolution of the homogeneity measure $\Phi$ in the random magnetic trap for the same parameters as in Fig.~\ref{fig:ncr_dynamo}:
\textbf{(a)}~homogeneous injection with re-injection and \textbf{(b)}~injection in a spherical shell
with re-injection.
The insets show the probability density of $\ncr/\mean{\ncr}$ at $t = 6857\, \rm yr$,
as in Fig.~\ref{fig:dan_stats},
with the parabolic and linear variations representing the log-normal and exponential distributions, respectively.
}
\label{fig:dynamo_stats}
\end{figure}

To confirm that the asymptotic value of $\Phi$ is unity for the homogeneous injection while $\Phi\neq1$ as $t\to\infty$ and/or $N\to\infty$ for injection in a spherical shell, we fitted the time variation of the computed values of $\Phi$ with the form $\Phi = c-d t^{-\gamma}$. 
The best-fit parameters $c$, $d$ and $\gamma$ are shown in Table~\ref{tab:conv_case1}. The quality of the fits is illustrated in Fig.~\ref{fig:phi_fit} where we plot $\log[(c-\Phi)/d]$ versus $\log t$. For a perfect fit, this is a straight line with the slope $-\gamma$, and the quality of the fit is remarkably good. It is notable that the fits are better when obtained excluding smaller values of $t$ where the number of particles involved is lower and short-term transients may not have died away yet. 
In the case of spherical shell injection into
the random field, convergence is slightly less monotonic, resulting in the relatively larger deviations from the fit at large $t$;
but the misfit remains very small.
It is reassuring that $\gamma$ increases as $N$ increases, so that $\Phi$  converges faster to the asymptotic values for larger $N$ at $t$ increases, as shown in Fig.~\ref{fig:danerrstatconvg}: $\Phi\approx1$ for the homogeneous injection and $\Phi\approx 0.8$ for the spherical-shell injection.
Although $d$ also increases as $N$ increases, the effect of this is outweighed by the increase in $\gamma$, so the convergence is not affected.
Table~\ref{tab:conv_case1} confirms that the number of particles in our simulations is sufficiently large to represent their distribution at large times, since the fit parameter $c$ varies little with $N$, and as noted above the combined variation in $\gamma$ and $d$ act to make convergence faster as $N$ increases.
We observe confidently the convergence of $\Phi$ to its asymptotic values.

We note that even in the case of homogeneous and isotropic particle injection when the Liouville theorem applies, it would be difficult to achieve the asymptotic state with $\Phi=1$ in any finite simulation because of the unavoidable inhomogeneities resulting from the finite spatial resolution.
Our fits to the dependence of $\Phi$ to $t$ and $N$ allow us to assess the asymptotic state that is realised.

\subsection{Protons in the isolated random trap}\label{TiaRF}

As discussed in Section~\ref{sec:magnetic_fields}, random magnetic fields have numerous magnetic traps \citep{Seta2018}. 
In Section~\ref{DGF}, we have isolated a region with one of the highest particle concentrations in a random magnetic field generated by the fluctuation dynamo to explore it in finer detail here. Despite the high efficiency of this trap, we have no reasons to consider it to be too unusual. This magnetic trap is a realization of a random magnetic field and is by far more realistic than the axisymmetric trap discussed above. Despite the complexity (it has three identifiable magnetic mirrors), the particle trajectories near the maxima of the magnetic field strength have the form typical of that near a magnetic mirror  \citep{Seta2018}.

The magnetic field in this trap is represented by a wide range of scales including those smaller than the Larmor radius of the particles in the simulation. 
This affects the particle magnetic moment which, however, varies little for long enough to allow multiple reflections before a particle escapes from the trap. 
On the other hand, the magnetic field varies over length scales large enough to trap particles within the energy range considered.

Fig.~\ref{fig:ncr_dynamo} presents the particle distribution in the random trap for the homogeneous injection with particle re-injection (Fig.~\ref{fig:ncr_dynamo}a) and for the injection in a spherical shell with re-injection in Fig.~\ref{fig:ncr_dynamo}b. The radii of the spherical shell were carefully chosen to ensure that the magnetic lines connect the injection region with the magnetic trap. 

As with the axisymmetric trap and in accordance with the Liouville theorem, the homogeneous injection with homogeneous re-injection of particles lost through the boundaries does not produce any systematic spatial variation in the particle number density. The re-injection affects profoundly
the particle distribution in space: the particles are still reflected at the magnetic mirrors and trapped.
 
Without the re-injection, when the particles are lost through the boundaries and their number decreases as the simulation progresses, the particle spatial distribution remains nonuniform at all times.
The Liouville theorem does not preclude this because the particle losses correspond to a sink term in the equation for the particle distribution function. As with the axisymmetric trap, the inhomogeneous injection (Fig.~\ref{fig:ncr_dynamo}b)  produces significant density variation.

Fig.~\ref{fig:dynamo_stats} shows the variation of the inhomogeneity parameter with time, similarly to Fig.~\ref{fig:dan_stats}. As in the axisymmetric trap, $\Phi\to1$ as $t$ increases in the case of homogeneous injection, whereas $\Phi$ tends
to a value of about $0.5$ for the spherical-shell injection. Also as in the axisymmetric trap, the probability density of $\ncr/\mean{\ncr}$ is close to log-normal in the former case. 
The probability density under the spherical-shell injection is different from that in the axisymmetric trap; 
although both have a similar structure, the case of the random trap has a pronounced exponential tail. 
We also note that the range of variation in $\ncr/\mean{\ncr}$ is much wider in the case of inhomogeneous injection confirming that it reflects systematic density variations in contrast to statistical fluctuations within a narrow range that occur when the particles are injected homogeneously; these features are common to the axisymmetric and random traps and appear to be generic.

We conclude that persistent inhomogeneity in the particle distribution develops when their injection is inhomogeneous (such as injection in a spherical shell) or the particles are not re-injected in the trapping region. This has been demonstrated with a simple, axisymmetric trap (Section~\ref{PDiAT}) and a complex trap in a realization of a random, intermittent magnetic field (Section~\ref{TiaRF}). As the number of particles and/or the duration of the simulation increase, the degree of inhomogeneity tends to $\Phi\approx0.8$ in the former and $\Phi\approx 0.4$ in the latter case.
\begin{figure}
    \centering
    \includegraphics[width=0.45\textwidth]{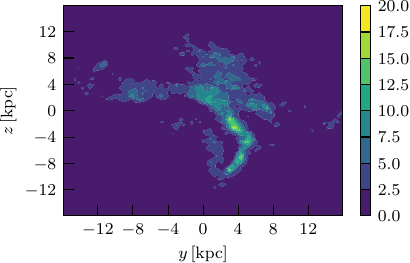}
    \caption{As in Fig.~\ref{fig:ncr_dynamo}b, in the cross-section of the random trap 
    integrated over $5\leq x \leq 6\,\rm pc$ at $t=6944 \,\rm yr$, but for injection from a 
    point source at the center of the box.
    }
    \label{fig:propoint}
\end{figure}

Fig.~\ref{fig:propoint} presents the
number density distribution in the isolated random trap for a different inhomogeneous source condition, 
whereby particles are injected from a point source at the center of the box. 
For $r\La  =1.2\,\rm pc$ the particles span significant distances, 
and in different cross sections we see multiple points of maxima.
Fig.~\ref{fig:propoint} shows significant maxima in the region $0\leq y \leq 5\,\rm pc$ and $-9\leq z \leq 1 \,\rm pc$. The structure of the inhomogeneity in this region is similar to the mirroring structure shown in Fig.~\ref{fig:n_e_random}a. 
This demonstrates that the conclusions drawn from the spherical shell case are consistent for other inhomogeneous source conditions,
which would however require larger number of particles evolved for longer times to obtain statistically significant signatures for the whole box.
\begin{figure*}
     \subfloat{
     \includegraphics[width=0.96\textwidth]{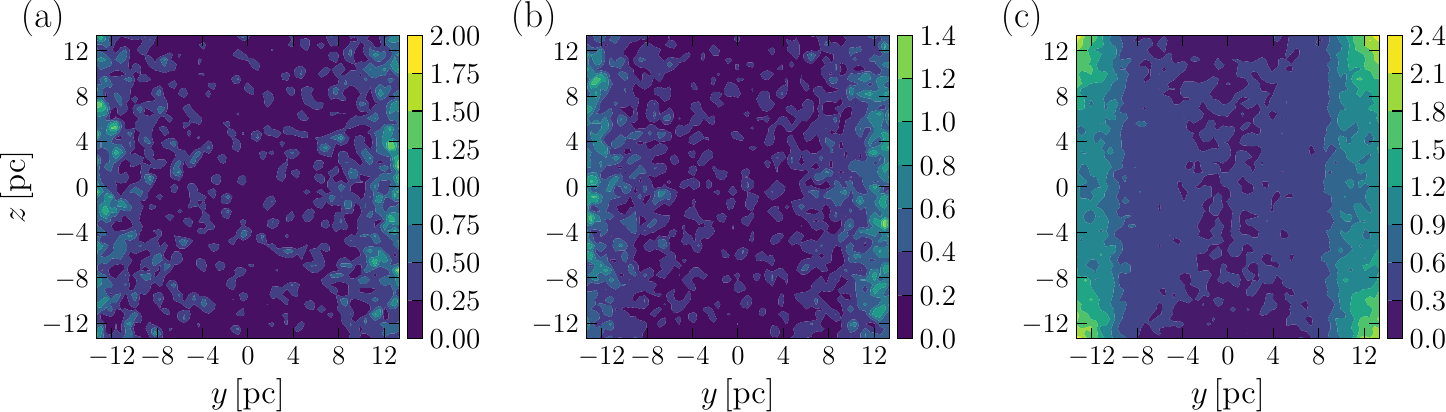}}\\
    \subfloat{
    \centering
         \includegraphics[width=0.95\textwidth]{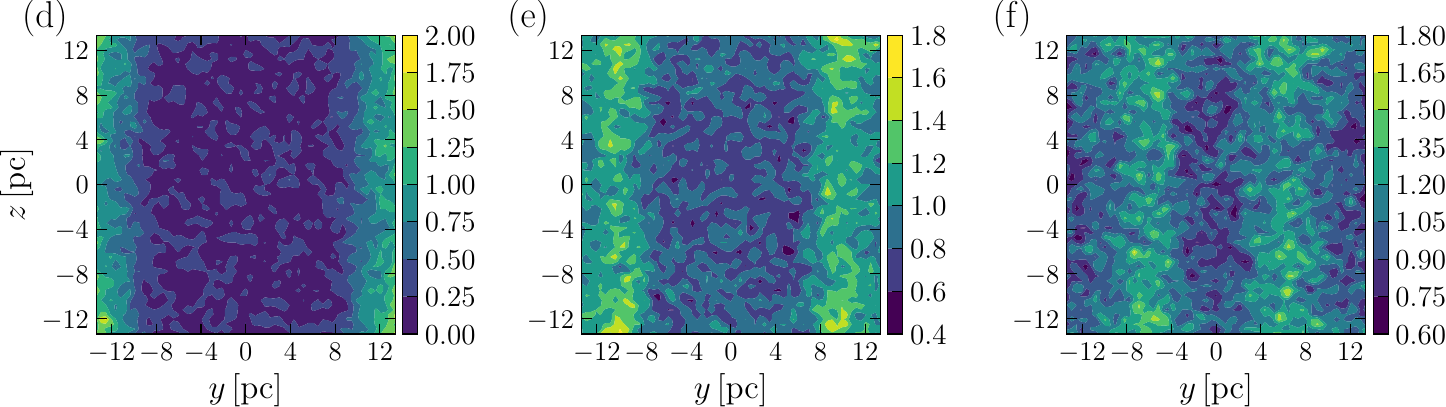}}\\
        \subfloat{
        \centering
         \includegraphics[width=0.95\textwidth]{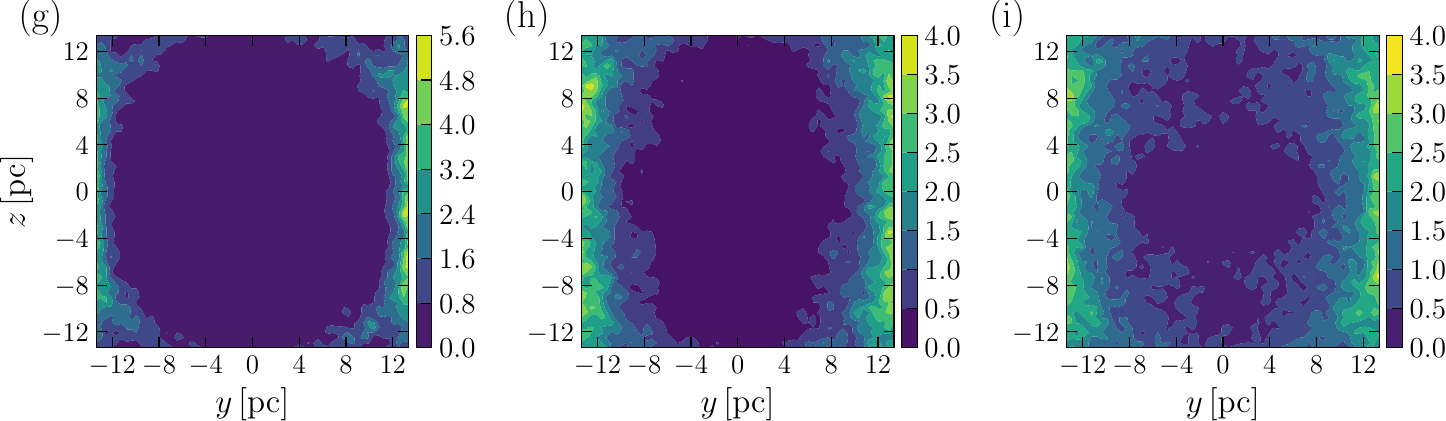}}
        \caption{The relative number density 
        $\ncr/\mean{\ncr}$
        (color-coded) of CR electrons of various energies in the cross-section of the axisymmetric trap at $x=0$ (integrated over $|x|\leq 6.4\,\rm pc$) at $t= 6944\,\rm yr$. In the upper row, the particles are all injected homogeneously (with re-injection) with the energy corresponding to the Larmor radius $r\La=0.62\,\rm pc$, and the distributions shown are for the particles with decreasing energies:
        \textbf{(a)} $r\La=0.62\,\rm pc$ , \textbf{(b)} $0.38\,\rm pc$, and \textbf{(c)} $0.19\,\rm pc$. Panels \textbf{(d)}--\textbf{(f)}, respectively, show similar distributions for the same particle energies
        and for the homogeneous injection,
        but for the case of the power-law energy injection spectrum $E^{-3/2}$. Panels \textbf{(g)}-\textbf{(i)} 
        show the case similar to
        that in Panels \textbf{(d)}--\textbf{(f)} but for the particles injected (and re-injected) in the spherical shell.
        }
                 \label{fig:randomdanloss}
\end{figure*}
\subsection{CR electrons: the effects of energy losses}
As noted above, when particles lost through the boundaries are not re-injected, the distribution of the protons is inhomogeneous even when they are injected homogeneously and isotropically.
Energy losses to the synchrotron radiation and inverse Compton scattering affect CR electrons of any given energy in a similar manner, so their spatial distribution is expected to be inhomogeneous even when they are injected continuously, homogeneously and isotropically.
Indeed, Fig.~\ref{fig:randomdanloss} shows the distributions of
the electrons of various energies (specified via the Larmor radius based on the r.m.s.\ magnetic field strength) in the axisymmetric trap. When all the particles are injected at the same energy,
a particle inhomogeneity develops in the region where the field strength is maximum and the energy loss is the strongest even when the particles are injected uniformly. In the lowest energy bin (Fig.~\ref{fig:randomdanloss}c), this produces the minimum in $\ncr/\mean{\ncr}$ around $x=0$ where the field is the strongest. The consequences of the energy losses are similar to 
those of an inhomogeneous injection (compare Figs~\ref{fig:randomdanloss}c and \ref{fig:ncr_dan_analytical015}b). However, the particle trapping is still evident as it produces local maxima in $\ncr/\mean{\ncr}$ along the $z$-axis at $y=0$.
Particles of lower energies (Fig.~\ref{fig:randomdanloss}b--c) partially fill the region around $y=0$: these are the particles lost from the energy bin shown in Fig.~\ref{fig:randomdanloss}(a). The minima in $\ncr/\mean{\ncr}$ at $y=0$ near the top and bottom of the frames are enhanced (in comparison with those in the proton distribution) by the stronger energy losses at those positions (where the magnetic field is stronger).
Fig.~\ref{fig:randomdanloss}(d--f) and \ref{fig:randomdanloss}(g--i) show the electron distributions for the case of the power-law injection energy spectrum for homogeneous  and spherical-shell injection regions, respectively. 
In Fig.~\ref{fig:randomdanloss}(d--f) we see maxima of number density being generated along the magnetic field lines defining the mirroring region.
As we move from higher to lower energy bins, the maxima shift towards the centre of the box, due to the change in trapping conditions as the Larmor radius decreases with energy, with the stronger magnetic fields in the central regions trapping the lower energy particles more efficiently.

\begin{figure*}
     \centering
    \subfloat{
         \includegraphics[width=0.98\textwidth]{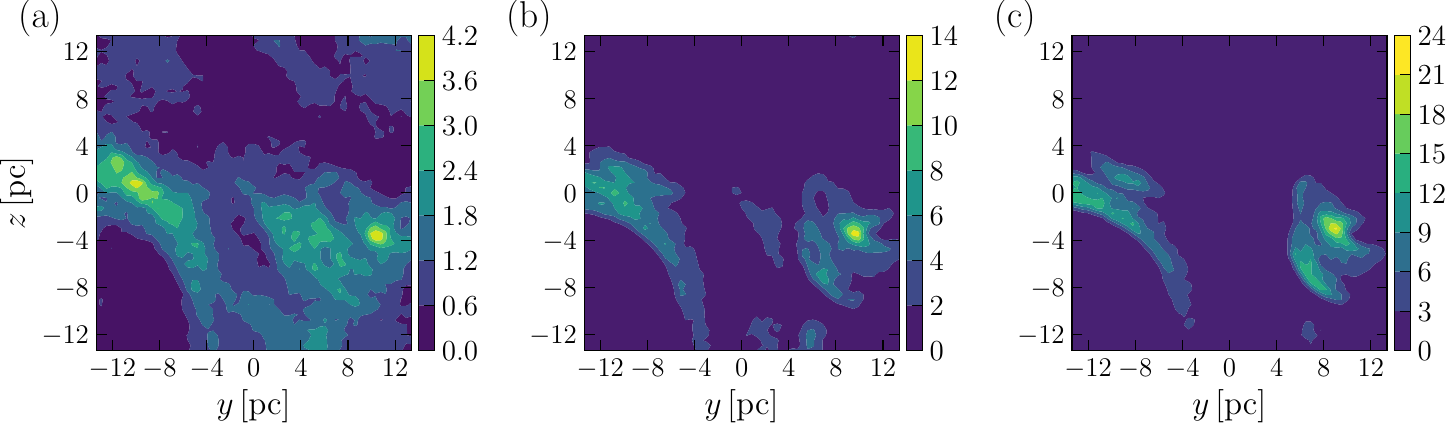}}\\
        \subfloat{
         \includegraphics[width=0.98\textwidth]{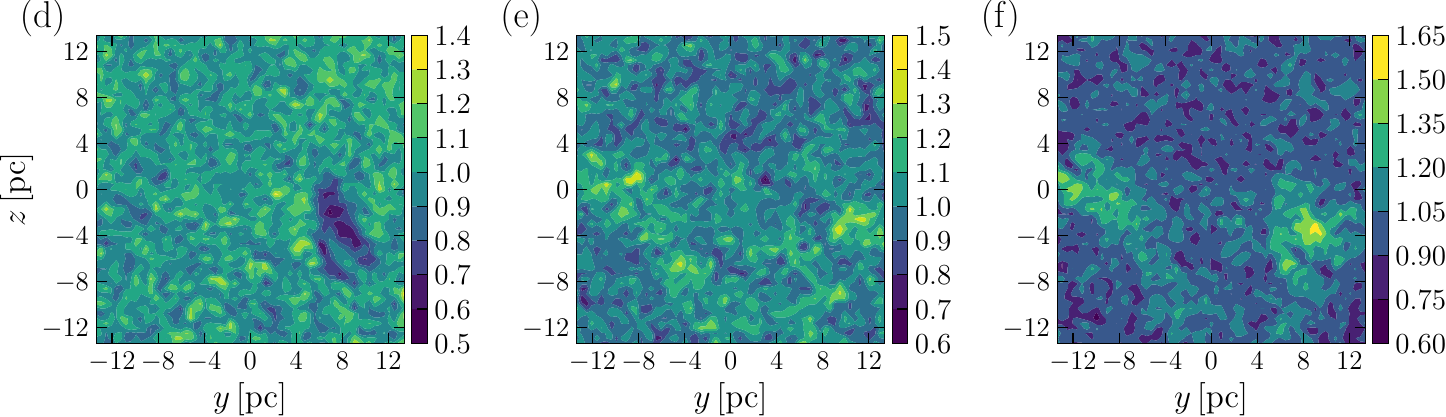}}\\
        \subfloat{
         \includegraphics[width=0.98\textwidth]{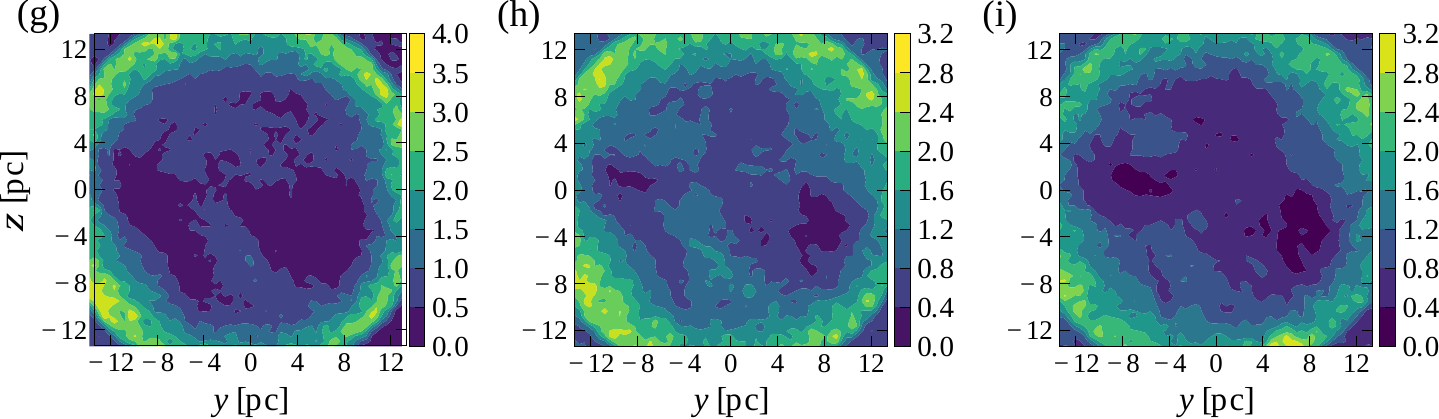}
         }
     \caption{
As in Fig.~\ref{fig:randomdanloss} but for the random trap with the electron distributions shown for the particles with decreasing energies, $r\La=0.69$, $0.45$ and $0.23\, \rm pc$ (based on the r.m.s.\ magnetic field strength, with the maximum energy on the left). 
The two upper rows show homogeneous injection (with re-injection) of particles with: 
\textbf{(a)--(c)} the same energy, and \textbf{(d)--(f)} the power-law injection energy spectrum $E^{-3/2}$. The lower row, \textbf{(g)--(i)}, shows injection in the spherical shell (with re-injection) with the power-law injection energy spectrum $E^{-3/2}$.
  }
     \label{fig:n_e_random}
\end{figure*}

\begin{figure}
    \centering
    \includegraphics[width=0.45\textwidth]{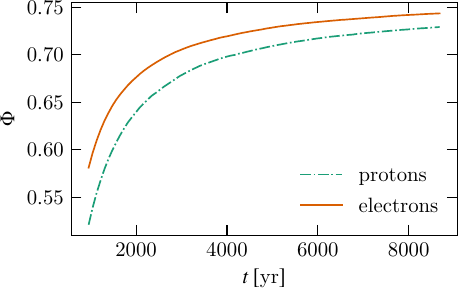}
    \caption{The evolution of the
    measure of inhomogeneity $\Phi$ for the electrons (solid) and protons(dash-dotted) injected into an isolated random trap from a spherical-shell source with the power-law energy spectrum  $E^{-3/2}$, 
    shown for the lowest-energy bin corresponding to $r\La = 0.23 \,\rm pc$.
    The particles are re-injected when they are lost through the boundaries or when they reach the minimum energy. (The difference in $\Phi$ for the protons from that in Fig.~\ref{fig:dynamo_stats} is due to the fact that here the particles are injected with the power-law energy spectrum.) }
    \label{fig:phi_electrons_protons_dynamo}
\end{figure}

The interplay between the particle trapping and energy losses is even more remarkable in the case of the random trap where the magnetic field strength has local maxima not only at magnetic mirrors but at other locations too. Similarly to Fig.~\ref{fig:randomdanloss}, Fig.~\ref{fig:n_e_random} shows the number density distribution for the electrons in various energy bins.
The number density for homogeneous injection with re-injection from a delta energy source (Fig.~\ref{fig:n_e_random} a-c), and for power-law injection(Fig.~\ref{fig:n_e_random} d-f),
have multiple maxima with two distinct regions in all energy bins, corresponding to the mirroring regions. 
For spherical shell injection with re-injection from a power-law source (Fig.~\ref{fig:n_e_random} g-i), 
the maxima are in slightly different locations, closer to the mirroring regions, 
because the source particles populate the magnetic field lines differently from the case of homogeneous injection (Fig.~\ref{fig:n_e_random} d-f).
As for the axisymmetric trap, inhomogeneities in the particle distribution become more pronounced at lower energies. When the electrons are injected with a power-law energy spectrum
with a negative spectral index
(Fig.~\ref{fig:n_e_random}d--f), most particles are injected at lower energies and, in the case of the homogeneous injection, this overwhelms the inhomogeneities that develop while the particles propagate. This distorts inhomogeneity measures such as $\Phi$, making them less informative. Therefore, we present $\Phi$ only for the case of the spherical-shell injection. Fig.~\ref{fig:phi_electrons_protons_dynamo} shows the evolution of $\Phi$ for the electrons and, for comparison, the protons injected within a spherical shell into the isolated random trap with the power-law energy spectrum $E^{-3/2}$. 
(The particular choice of the energy spectra, which is flatter than the more commonly-adopted $s=-2$, is to ensure that the number of particles in the higher energy bins remains statistically significant to study the evolution of number densities. From test runs the $s=-2$ case shows similar trends in lower energy bins, with enhanced loss rate for particles in the higher energy bins.
The results discussed hereafter focus on features with conservatively well-resolved density distributions.)
In this case, the particles are re-injected when they are lost through the boundaries of the computational domain or when they reach the minimum energy. The spatial distributions of both the protons and 
electrons are significantly inhomogeneous, with $\Phi$ remaining smaller than unity, asymptotically $\Phi\approx 0.75$ for both protons and electrons. 

\begin{table}
\caption{The cross-correlation coefficients between the magnetic field energy density $B^2$ and the CR proton and electron number densities, $n_\text{p}$ and $n_\text{e}$, respectively, the latter at the lowest energy corresponding to $r\La=0.16\,\rm pc$, in the random magnetic trap of Section~\ref{DGF} (the upper part of the table) and the full realization of the random magnetic field (Section~\ref{PDiRMF}, the lower part), for the two forms of the injection region, homogeneous (left) and injection in a spherical shell (right). The $1\sigma$ errors of the cross-correlation coefficients are of order $10^{-3}$ or less.}
    \centering
\begin{ruledtabular}
\begin{tabular}{@{}lccccccc@{}}
    &\multicolumn{3}{c}{Homogeneous}
       & &\multicolumn{3}{c}{Spherical shell}\\ 
        \cline{2-4} \cline{6-8}
&\multicolumn{7}{c}{Isolated random trap}\\
                &$B^2$                      &$n_\text{p}$    &$n_\text{e}$    
                                                    &&$B^2$             &$n_\text{p}$    &$n_\text{e}$\\
       \cline{2-4} \cline{6-8}
$B^2$           &1      &\llap{$-$}0.02     &0.001              
                                                    &&1     &\llap{$-$}0.01     & 0.12\\
$n_\text{p}$ &\llap{$-$}0.02\phantom{0}  &1                  &\llap{$-$}0.01\phantom{0}
                                                    &&\llap{$-$}0.01    &1                  &0.86\\
$n_\text{e}$ &0.001                      &\llap{$-$}0.01    &1
                                                    &&0.12              &0.86               &1\\
&\multicolumn{7}{c}{Random magnetic field}\\
                &$B^2$      &$n_\text{p}$    &$n_\text{e}$    &&$B^2$ &$n_\text{p}$    &$n_\text{e}$\\
        \cline{2-4} \cline{6-8}
$B^2$           &1          &0.003              &\llap{$-$}0.004    &&1     &0.001          &0\\
$n_\text{p}$ &\llap{$-$}0.003   &1                  &0.005          &&0 &1      &0.93\\
$n_\text{e}$ &\llap{$-$}0.004   &0.005              &1              &&0.001     &0.93\phantom{0}           &1\\
\end{tabular}
\end{ruledtabular}
    \label{CC}
\end{table}

\begin{figure}
    \centering
    \includegraphics[width=0.45\textwidth]{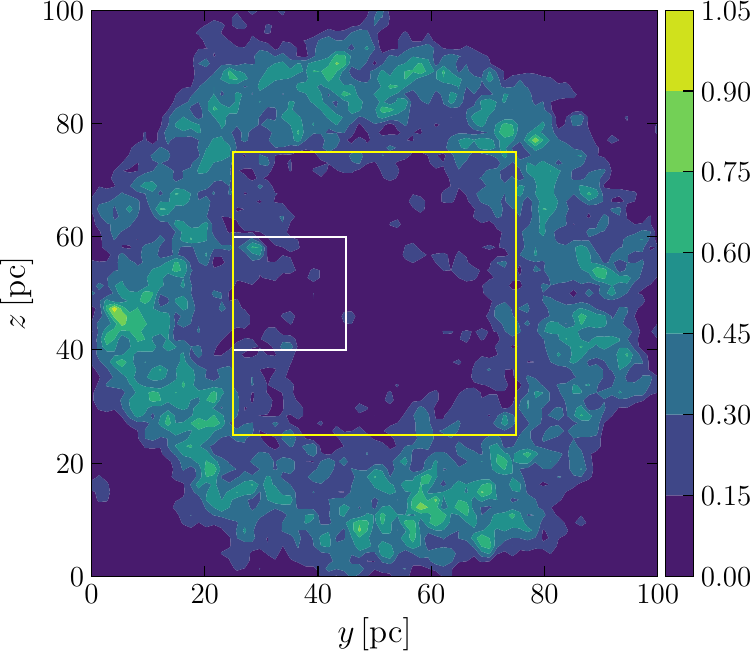}
    \caption{The relative number density (color-coded) of CR electrons of various energies in the cross-section of the full realization of the random magnetic field at $x$ integrated over $20 \leq x\leq 32\,\rm pc$ at $t= 6944\,\rm yr$ for the inhomogeneous injection with re-injection (the form of the injection shell is visible as the annular
    maximum in the particle density). The particles are injected with the energy drawn from the power-law energy spectrum $E^{-3/2}$, and the Larmor radius  corresponding to the energy bin shown here is $r\La=0.45\,\rm pc$. The yellow frame shows the position of the region shown in Fig.~\ref{BPE}.}
    \label{fig:dynamo_fullbox_electrons_shell}
\end{figure}

\begin{figure*}
\centering
\includegraphics[width=0.9\textwidth]{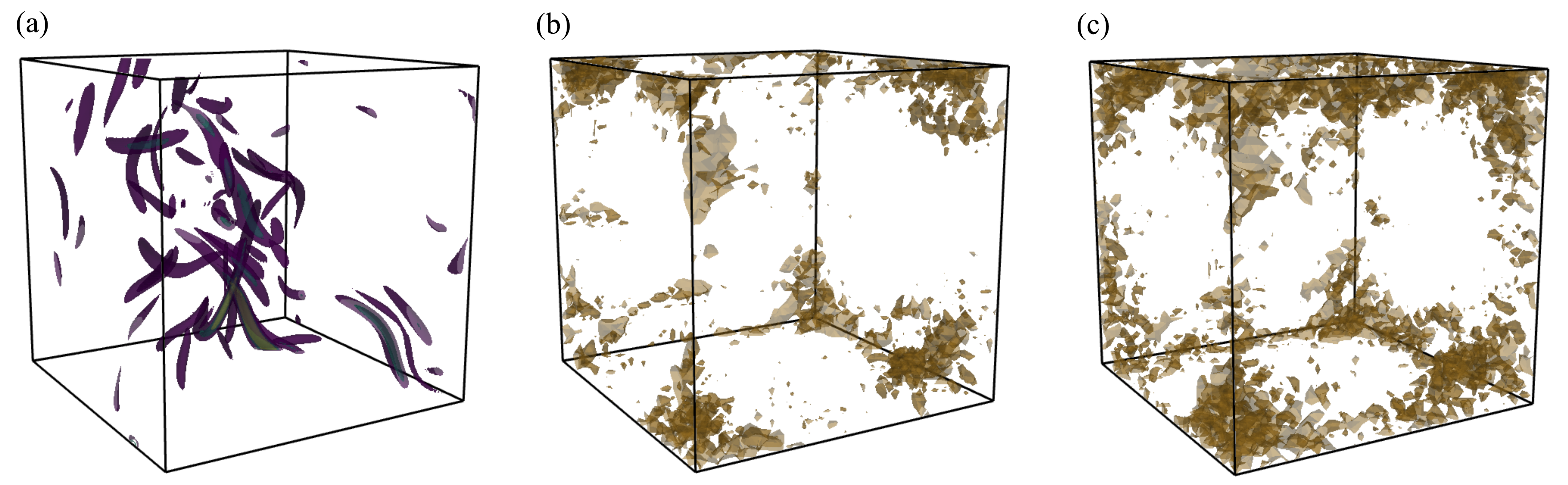}
\caption{The isosurfaces of \textbf{(a)}~the magnetic field strength, at $B/B_\text{rms}= 3.5$ and 6.5, and the time-integrated (for  $t\leq6344 \,  \rm yr$) number densities of \textbf{(b)}~protons, at $n_\text{p}/\langle n_\text{p} \rangle= 2$ and 8, and \textbf{(c)}~electrons, at $n_\text{e}/\langle n_\text{e} \rangle =2$ and $8$,
in the full realization of the random magnetic field of Section~\ref{PDiRMF}. The particles are injected in a spherical shell at the periphery of the domain (visible in Fig.~\ref{fig:dynamo_fullbox_electrons_shell} as the circular region of a high number density). To avoid the overcrowding of the isosurfaces, this figure shows the inner part of the domain $27\, {\rm pc} \leq (x,y,z) \leq 77\, \rm pc$ indicated with the yellow frame in Fig.~\ref{fig:dynamo_fullbox_electrons_shell}. The particle concentrations in the corners are a signature of the particle source. Despite the overall similarity in the distributions of the protons and electrons, they are clearly distinct from each other and from the
spatial distribution of the
magnetic field strength.}\label{BPE}
\end{figure*}

\subsection{Particle distribution in a full realization of a random magnetic field}\label{PDiRMF}
As discussed above,
a generic random magnetic field (either Gaussian or intermittent) contains numerous magnetic traps, one of which is described in Section~\ref{DGF}. In this section we discuss the distributions of the CR protons and electrons in a realization of the intermittent random magnetic field generated by the fluctuation dynamo, of which the trap of Section~\ref{DGF} is a small part. The particle distributions in such a magnetic field depend on the details of the injection and re-injection (such as the shape of the injection region) and energy losses in a manner similar to the case of isolated magnetic traps discussed above. For the homogeneous injection, we use periodic boundary conditions which essentially act as re-injection. In the case of the inhomogeneous particle source, we re-inject the particles
in a spherical shell when they are lost through the boundaries or because of energy losses.
The boundaries of the injection region in this case 
are fixed as $r_{\rm min} = 24 \, {\rm pc}$ and $ r_{\rm max} = 50 \, {\rm pc}$, 
with the full length of the periodic domain being 100\,pc. 

Fig.~\ref{fig:dynamo_fullbox_electrons_shell} shows the  distribution of electrons injected from the inhomogeneous source (spherical shell) with the $E^{-2}$ energy spectrum at the injection. The variation of $\Phi$ with time confirms the persistence of numerous local inhomogeneities. As with the isolated traps, the particle distribution becomes increasingly homogeneous with time for both electrons and protons injected from a homogeneous source, similarly to the cases shown in Panels~(d)--(f) of Figs~\ref{fig:randomdanloss} and \ref{fig:n_e_random}.

Fig.~\ref{BPE} shows the local maxima of $B$, $n_\text{p}$ and $n_\text{e}$ in a part of the domain (as indicated by the white frame within Fig.~\ref{fig:dynamo_fullbox_electrons_shell}). The maxima in the particle distributions in the corners of the cube shown are due to particle diffusion from the spherical-shell injection region. It is notable that the maxima in the particle number density are unrelated to those of the magnetic field strength, as quantified by their low cross-correlation coefficients in Table~\ref{CC}. Moreover, the proton (Panel b) and electron (Panel c) distributions are rather different, especially regarding their higher maxima, even though their cross-correlation coefficient (discussed below) is significant. The maxima in the electron distribution at the energy shown ($r\La=0.23\,\rm pc$), 
which are at different locations from those for the protons, 
are populated by particles that have lost some of their energy earlier and thus responded to different features in the magnetic field than the protons.
The energy ranges we explore, and the energy loss coefficient $\kappa$ we use, 
cause the particle to completely lose its energy within the time scales of the simulation. This makes it difficult to explore the evolution of spectra in energy space. Theoretical models of the evolution of spectra in uniform magnetic fields show a spectral steepening
by a power of 1.
We do not see any such spectral steepening in our simulations. Our understanding is that while our simulation setup is suitable for exploring the spatial distribution of cosmic ray particles, 
the numerical limitations in choosing the minimum energy and the resolution in the energy space 
do not allow  us to study the full scope of the spectral evolution.
\section{Relative distributions of protons, electrons and magnetic field}

Many interpretations of the radio-astronomical observations of galactic and extragalactic magnetic fields and cosmic rays rely on the assumptions that CR and magnetic field energy densities are equal (or proportional) to each other and, in addition, that the CR electrons (which produce the synchrotron emission observed) and protons (which dominate the CR energy density) have identical spatial distributions \citep[see e.g.\ section 4.5 of Ref.][]{ShSu21}.
Our results suggest strongly that both assumptions are not justified \citep[see also Ref.][]{Seta2018}.
The relation between the distributions of the particles and magnetic field can be characterized with the cross-correlation coefficient for the constituents $\mathcal{A}$ and $\mathcal{B}$,
\begin{equation}
    C(\mathcal{A},\mathcal{B}) = \frac{ \mean{\mathcal{A}\mathcal{B}} - \mean{\mathcal{A}} \, \mean{\mathcal{B}}}{\sigma_\mathcal{A} \sigma_\mathcal{B}},
\end{equation}
where angular brackets denote the spatial averaging and $\sigma_\mathcal{A}$ and $\sigma_\mathcal{B}$ are the standard deviations of $\mathcal{A}$ and $\mathcal{B}$, respectively (e.g., $\sigma_\mathcal{A}^2=\mean{\mathcal{A}^2}-\mean{\mathcal{A}}^2$).
The cross-correlation coefficients are shown in Table~\ref{CC} for both the single trap taken from a random magnetic field (Section~\ref{DGF}) and the whole realization of that field (Section~\ref{PDiRMF}). 
The results presented are for the electrons of the lowest energy but they vary little with the particle energy.

\begin{figure*}
    \centering
    \includegraphics[width=0.8\textwidth]{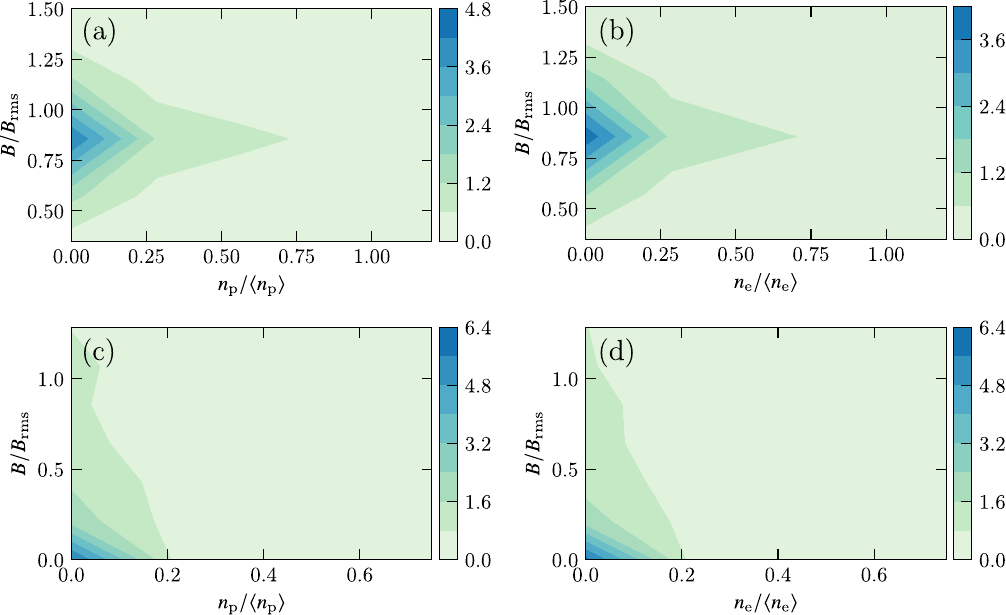}
    \caption{The joint probability density of the particle number density and the magnetic field strength in a random magnetic field: with homogeneous injection for \textbf{(a)} protons and \textbf{(b)}  electrons; and with spherical shell injection for \textbf{(c)} protons and \textbf{(d)} electrons.
    }
    \label{fig:joint_probability}
\end{figure*}

The distributions of the CR particles are uncorrelated with the magnetic field strength, irrespective of the choice of the injection region, and despite the fact that the electrons are sensitive to the magnetic field strength through their energy losses. The distributions of the CR protons and electrons are uncorrelated in the case of homogeneous injection (when both distributions are nearly uniform) but exhibit 
a significant correlation when injected inhomogeneously (in a spherical shell, when their inhomogeneities are persistent). The correlation coefficient of the proton and electron distributions in this case slightly increases with the total number of particles involved in the simulation.

Fig.~\ref{fig:joint_probability} shows the joint probability distributions of the particle number density
and magnetic field strength for the case of the full realization of the random magnetic field discussed in Section~\ref{PDiRMF}. 
The form of the joint probability distributions is sensitive to the form of the injection region. For a homogeneous source the particles tend to be localized in regions where the magnetic field strength is close to its root-mean-square value; these are just typical regions in the domain and this tendency does not suggest any causal connections between the magnetic field and particle distributions. The situation is different in the case of the spherical-shell injection shown in Panels (c)--(d): here the particles tend to stay in regions of weaker magnetic field.

The spatial distributions of the CR electrons and protons are not only uncorrelated with the magnetic field strength, but even more notably, the particle and magnetic field strength spatial
distributions are statistically independent, for both homogeneous and inhomogeneous particle injections. That is, the joint probability density $p(B,\ncr)$ (with $\ncr$ for either protons or electrons) is close to the product of the individual probability densities, $p(B)$ and $p(\ncr)$. We demonstrate this using the diagnostic
\begin{equation}\label{X}
X = \frac{\int\left[p(B,\ncr)-p(B)p(\ncr) \right]^2\,\dd B\,\dd n}{\int p^2(B,\ncr)\,\dd B\,\dd n}\,,
\end{equation}
where the integration extends over all values of the variable available (in practice, we use binned data and the integrals reduce to sums). For statistically independent variables, $X=0$. The probability density distributions for $\ncr/\langle \ncr \rangle$ (both the protons and electrons) and magnetic field $B/\langle B_{\rm rms} \rangle$ were calculated for the values of these ratios in the range from $0.005$ to $3.5$ and collected into ten bins of equal widths. 
Both the proton and electron distributions in the full realization of the random magnetic field (e.g., Fig.~\ref{fig:dynamo_fullbox_electrons_shell}) have $X\approx 10^{-4}$ for homogeneous injection and $10^{-3}$ when the particles are injected in the spherical shell. 
This indicates that both particle distributions are very close to being statistically independent of the magnetic field. 
We note that statistical tests for the statistical independence are only efficient for continuous random variables, and their usefulness is affected by data binning, which is unavoidable in the case of numerical results obtained at finite spatial and temporal resolutions \citep{Hoeffding1994}.
The arguments for the statistical independence provided above are subject to the same limitations but they are simple, transparent and remain stable when the data are binned differently.

To summarize, the spatial distributions of the CR particles are not only uncorrelated but even statistically independent of the magnetic field strength for both protons and electrons. This is the consequence of the fact that the distribution of particles with a relatively small Larmor radius is controlled not by the field \textit{strength} but by its \textit{structure}, in particular by magnetic traps which can occur in either weak or strong magnetic field regions. This conclusion is even more striking in the case of the CR electrons since their distribution is indeed affected by the local magnetic field strength because of their energy losses to synchrotron emission. And yet, the magnetic mirroring appears to dominate the distribution of the electrons.

\subsection{Implications for synchrotron intensity}\label{IfSI}
The intensity of the synchrotron emission $I$ of an astronomical object depends on the number density of cosmic-ray electrons $n_\text{e}$ and the magnetic field strength $B_\perp$ in the plane perpendicular to the line of sight $\vec{s}$ \citep[see, e.g., Ref.][for details]{SS21}
\begin{equation}\label{Iorig}
    I  \propto \int_{-\infty}^{L} n_\text{e} B_\perp^\alpha\,\dd s \,,
\end{equation}
where $\alpha$ depends on the energy spectrum of the CR electrons, $s$ is the position along the line of sight with $s=L$ at the observer, and $\alpha=2$ can be adopted as a reasonable approximation. 
In practice, Eq.~\eqref{Iorig} involves volume integration over the beam cylinder.
The integral can be identified with the spatial average which, using the ergodic assumption, can be replaced with the ensemble average, leading to 
\begin{align}
I &\propto
\iint_{-\infty}^\infty 
p (n_\text{e},B_\perp^2)\,\dd n_\text{e}\,\dd B_\perp^2\nonumber\\
&=\int_{-\infty}^\infty 
 p(n_\text{e}) \,\dd n_\text{e} 
        \int_{-\infty}^{\infty} 
p(B_\perp^2)\,\dd B_\perp^2\,,
\end{align}
where $ p(n_\text{e},B_\perp^2)$ is the joint probability distribution of the CR electron number density and $B_\perp^2$ while $ p(n_\text{e})$ and $p(B_\perp^2)$ are the corresponding marginal distributions, and the second equality follows from the statistical independence of $n_\text{e}$ and $B^2$ established above. 
The statistical independence of $n_\text{e}$ and $B$ extends to $n_\text{e}$ and the powers of $B_\perp$.
Reverting back to spatial averages, the synchrotron intensity reduces to the product of two line-of sight integrals,
\begin{equation}\label{Isplit}
    I \propto \int_{-\infty}^{L} n_\text{e}\,\dd s \int_{-\infty}^{L}B_\perp^2\,\dd s\,.
\end{equation}
Each of these integrals is arguably easier to estimate or constrain than the original integral, which can facilitate significantly the interpretation of radio astronomical observations in terms of the CR and magnetic field properties.

We have demonstrated that $n_\text{e}$ and $B^2$ are statistically independent in the case of a random magnetic field. Correspondingly, the splitting of the integral \eqref{Iorig} into the product of two simpler integrals, as in Eq.~\eqref{Isplit}, is possible at those spatial scales and particle energies where the CR electron Larmor radius $r\La$ is comparable to the magnetic field scale, i.e., mostly at the turbulent scales in the interstellar medium. In particular, this implies that the widely used assumption of the local, point-wise equipartition between cosmic-ray and magnetic energy densities \citep{Seta2019} is inapplicable at those scales.

An important factor, inaccessible with the test-particle simulations used above,
is the possibility that the CR pressure drives plasma motions that modify the magnetic field. This can introduce connections between the CR and magnetic field distributions and statistical properties. This aspect of the CR propagation is largely unexplored. Another effect that can produce such connections is the large-scale dynamics of the interstellar medium such as the Parker instability.
However, simulations of the saturated states of the Parker instability have not revealed such a correlation \citep{D22}.
From their analysis of the synchrotron fluctuations at a scale of order $100\,\rm pc$ in spiral galaxies, \citet{SSFBLPT14} also suggest that the equipartition assumption is inapplicable. These authors also find that the distributions of the CR electrons and magnetic field strength are slightly anti-correlated at those scales. Such a connection might be a consequence of the dynamical effects of cosmic rays on the interstellar magnetic fields.

\section{\label{conclusion}Conclusions}

Through test particle simulations, we have demonstrated that CR particle trapping between magnetic mirrors, which are abundant in random magnetic fields, leads to persistent inhomogenities in the spatial distributions of both CR protons and electrons when the particles are injected inhomogeneously (but isotropically) and due to energy losses.
When the particles are injected uniformly and isotropically, the Liouville theorem precludes any significant, persistent inhomogeneities in the particle distribution provided, in particular, that the particle Larmor radius exceeds the scales at which the magnetic field varies (which may or may not be true in the interstellar medium, depending on the particle energy). 

There are several consequences of the particle trapping, beyond their inhomogeneous spatial distribution. In particular, the probability distribution of the particle number density $\ncr$ is close to being lognormal when the particles are injected homogeneously and isotropically (and thus their spatial distribution remains homogeneous), whereas it is more complicated (developing an exponential tail at large values of $\ncr$; see Figs~\ref{fig:dan_stats} and \ref{fig:dynamo_stats}) when the effect of the particle trapping on their spatial distribution is significant. The number densities of either protons or electrons are uncorrelated with the magnetic field strength $B$. Moreover the particle number densities are statistically independent of $B$ for both homogeneous and inhomogeneous injections. The particle distributions in space are controlled not by the strength of the magnetic field but rather by its structures, in particular, by magnetic traps where $\ncr$ can be larger between magnetic mirrors, where the magnetic field is weaker. 

The spatial distributions of protons and electrons are mutually correlated when they are injected inhomogeneously. The particle trapping and its effect on the distribution of CR particles with respect to the magnetic field distribution has implications for the interpretation of observations of synchrotron emission produced by CR electrons in random magnetic fields discussed in Section~\ref{IfSI}. We find no evidence to support the assumption of the equipartition between cosmic rays and magnetic energy densities at the turbulent scales. 

We believe that our conclusions, based on  simulations of particles of relatively high energies (due to numerical constraints on the spatial and temporal resolution of the simulations), also apply to particles with energy of the order of 1\,GeV that dominate the observable synchrotron emission of galaxies. Particle mirroring cannot be captured  by the standard fluid descriptions of cosmic rays used in magnetohydrodynamic 
simulations of the interstellar medium. 
These kinetic effects remain to be included into cosmic ray propagation models at turbulent scales.

\begin{acknowledgments}
We are grateful to Daniel Elsender for his contribution to the exploration of the magnetic trap of Section~\ref{subsec:danmirror}. Useful discussions with Torsten En{\ss}lin and Amit Seta are gratefully acknowledged. We also thank Luiz Felippe S.\ Rodrigues for  his help in code development and discussions. 
Insightful comments and suggestions from two anonymous referees are gratefully acknowledged.
\end{acknowledgments}

\appendix

\section{Sampling of particle trajectories and number density estimates}\label{aADF}

\begin{figure}
    \centering
    \includegraphics[width=0.9\columnwidth]{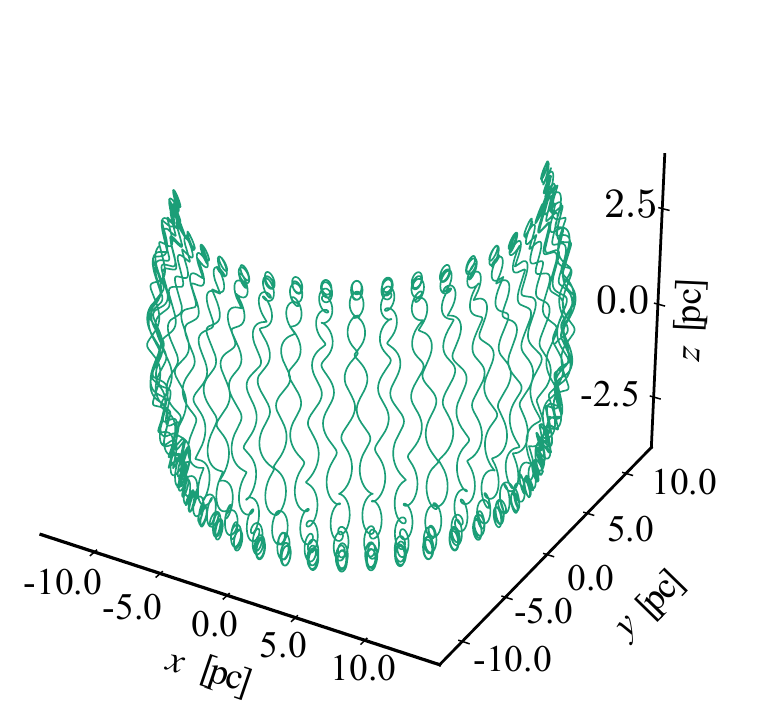}   
   \caption{The trajectory of a particle  of the Larmor radius  $r\La = 0.8 \, \rm pc$ (based on the r.m.s.\ magnetic field strength of $5\,\upmu\rm G$) bouncing between the 
    mirrors located near the poles of a magnetic point dipole.
   }
    \label{dipole}
\end{figure}

In addition to the examples of magnetic traps presented in Sections~\ref{subsec:danmirror} and \ref{DGF}, we also considered particle trapping in a point-dipole magnetic field aligned with the $z$-axis, given in Cartesian coordinates by
\begin{equation}
\begin{pmatrix} B_x\\ B_y\\ B_z\end{pmatrix}
=
\frac{3M}{r^5}\begin{pmatrix}xz\\ yz\\ z^2-\tfrac13 r^2 \end{pmatrix},
\end{equation}
where $r$ is the spherical radius and $M$ is the dipole moment. 
The magnetic field is defined in a cubic region of dimensionless edge length $2\pi$, and the particles are injected either at random positions uniformly distributed throughout the region, or in the spherical shell, as discussed in Section~\ref{sec:Numericalsetup}. 
We consider particles of a constant energy corresponding to the Larmor radius $r\La = 0.8 \, \rm pc$ (with the dimensionless length $2\pi$ corresponding to $ 100\,\rm pc$)
for the magnetic field strength of $3.36 \,\upmu\rm G$ at the equator,
$(z,r)=(0,5.35) \, \rm pc$. The typical particle trajectory is shown in Fig.~\ref{dipole}. The particles gyrate around the field lines, 
drift in
azimuth $\phi$,
and are reflected near the poles where the magnetic field satisfies the mirroring conditions for their energy and pitch angle. This example, where the intuitive expectations for the particle distribution are available, is useful to identify --- and avoid --- a bias in the particle number density estimates associated with the particle injection algorithms. 

\begin{figure*}
    \includegraphics[width=\textwidth]{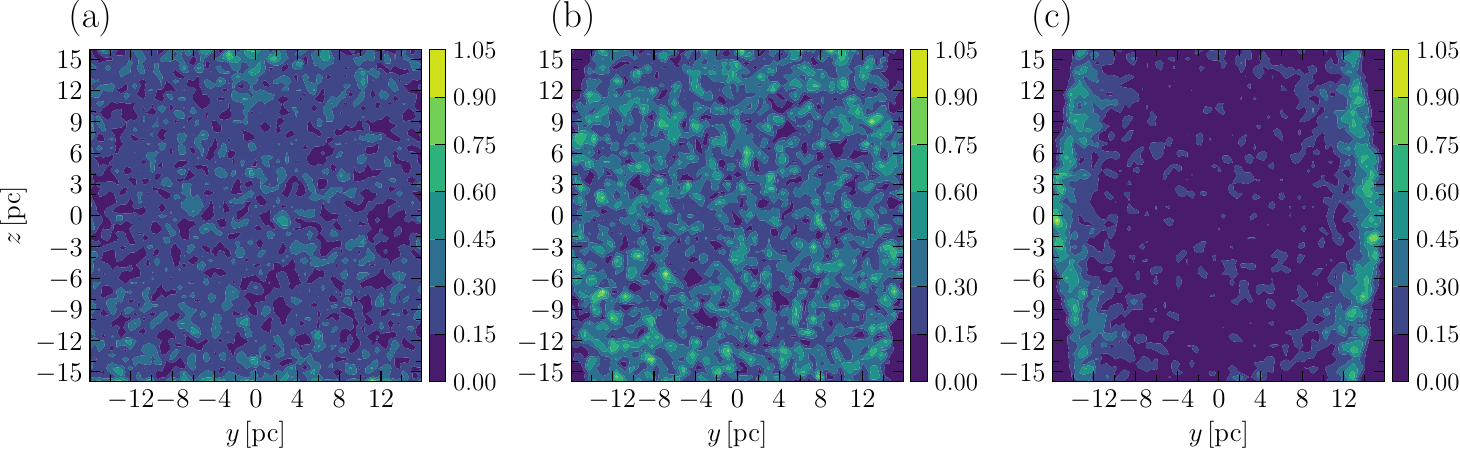}
    \caption{The relative number density $\ncr/\mean{\ncr}$ (color-coded) of CR protons in the cross-section of the dipole trap at $x=0$ (integrated over $|x|\leq 6.4\,\rm pc$) at $t= 200 \,\rm yr$ for
    \textbf{(a)}~homogeneous injection with reflective boundary condition,  \textbf{(b)}~homogeneous injection with re-injection and \textbf{(c)}~injection and re-injection in a spherical shell. The particle Larmor radius is $r\La=0.8\,\rm pc$.
    }
    \label{fig:ncr_dipole_analytical_015}
\end{figure*}

Fig.~\ref{fig:ncr_dipole_analytical_015} illustrates the particle distribution in the dipole trap for various injection schemes. The difference is significant and, apart from the difference in the injection algorithm, 
the fact that the particle number density is obtained by sampling particle trajectories as described in Section~\ref{NDoaEoP} contributes to it:
longer trajectories contribute more to $\ncr$ than the shorter ones. 

\begin{figure*}
    \centering
    \includegraphics[width=0.8\textwidth]{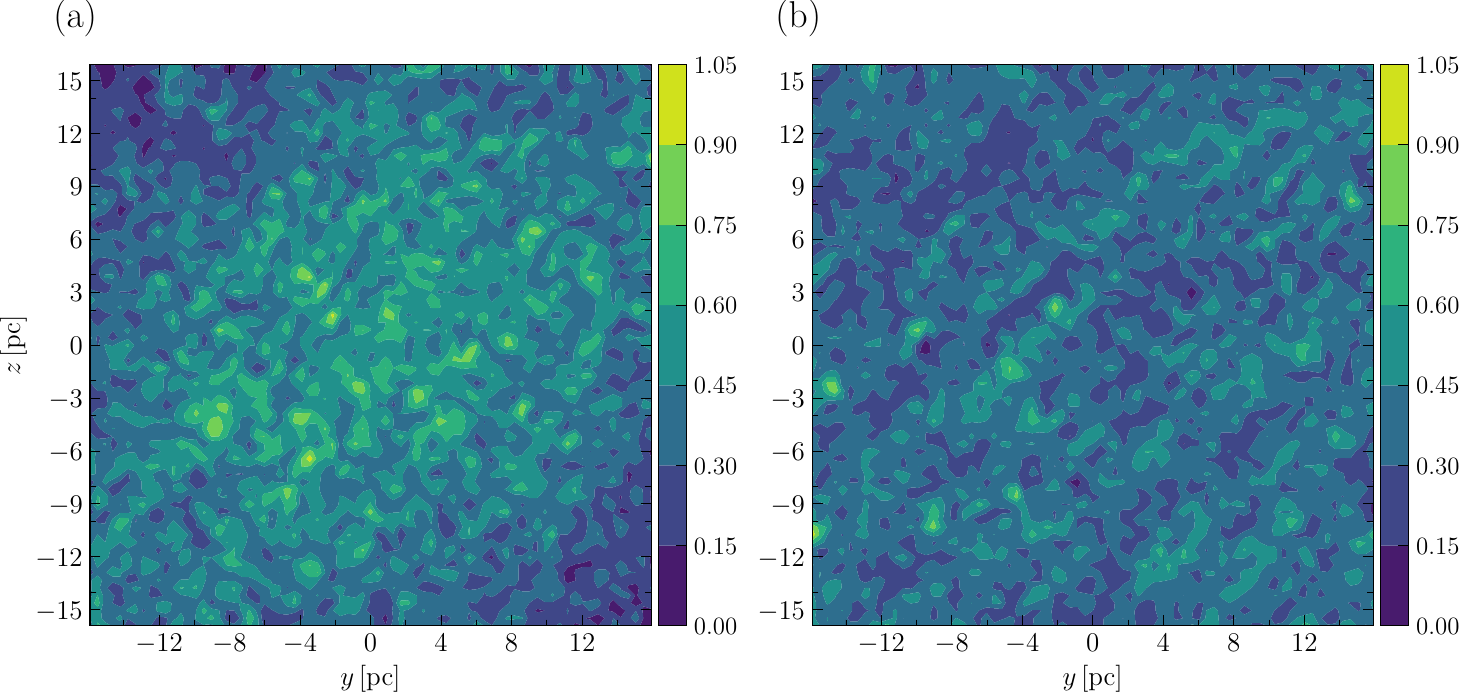}
    \caption{The relative number density of particles in a uniform magnetic field \eqref{eq:fielddiag} in the case of homogeneous injection with 
    \textbf{(a)}~homogeneous re-injection of particles lost through the boundaries of the computational domain and \textbf{(b)}~reflecting boundary conditions. The sampling bias produces spurious inhomogeneity in the particle distribution in Panel~(a).}
    \label{fig:diagonalfield}
\end{figure*}

The sampling bias is best understood in the case of a uniform, inclined magnetic field in a cubic region, e.g., 
\begin{equation}\label{eq:fielddiag}
    B_x=B_y=B_z\,,
\end{equation}
where the field lines through the central region are longer than those in the corners which are not connected by magnetic lines. 
Since the particle trajectories follow the field lines, the central locations are visited by particles injected uniformly at a larger set of locations. With particle losses through the region boundaries and their re-injection at random positions, the central regions therefore have a higher measured density. 
Fig.~\ref{fig:diagonalfield} shows the distribution of particles uniformly injected and re-injected into the magnetic field \eqref{eq:fielddiag}.

To avoid the bias, we use reflecting boundary conditions whereby the particle velocity is reversed in direction to bring the particle back into the region as it reaches the face of the cubic computational domain. As shown in Fig.~\ref{fig:diagonalfield}, this results in an appropriately uniform particle distribution. The bias is much less significant in random magnetic fields where all or most magnetic lines have similar lengths spanning the simulation domain and there is more freedom in the choice of the boundary conditions.

%

\end{document}